\documentclass[floatfix,superscriptaddress,numerical,amsmath,amssymb, pre, aps, reprint,showpacs]{revtex4-2}

\usepackage{graphicx}
\usepackage{caption}
\usepackage{hyperref}
\usepackage{paralist}
\usepackage{amsmath}
\usepackage{amssymb}
\usepackage{mathtools}

\usepackage{color}
\usepackage{verbatim}
\usepackage{comment,xcolor}
\newcommand{\mauro}[1]{\textbf{\textcolor{red}{[M:~#1]}}}

\usepackage[normalem]{ulem}

\DeclarePairedDelimiter\ceil{\lceil}{\rceil}
\DeclarePairedDelimiter\floor{\lfloor}{\rfloor}
\newcommand{\dtriangle}{\mathbin{\raisebox{2pt}{\protect\rotatebox[origin=c]{180}{$\triangle$}}}}

\begin{document}
\title{How does homophily shape the topology of a dynamic network?}
\author{Xiang Li}
\affiliation{Department of Applied Mathematics, School of Mathematics, University of
Leeds, Leeds LS2 9JT, United Kingdom}
\author{Mauro Mobilia}
\affiliation{Department of Applied Mathematics, School of Mathematics, University of
Leeds, Leeds LS2 9JT, United Kingdom}

\author{Alastair M. Rucklidge}
\affiliation{Department of Applied Mathematics, School of Mathematics, University of
Leeds, Leeds LS2 9JT, United Kingdom}

\author{R.K.P. Zia}
\affiliation{Center for Soft Matter and Biological Physics, Department of
Physics, Virginia Polytechnic Institute \& State University, Blacksburg, Virginia
24061, USA}
\affiliation{Department of Physics \& Astronomy, University of North Carolina at Asheville, Asheville, North Carolina 28804, USA}
\affiliation{Physics Department, University of Houston, Houston, Texas 77204, USA
}

\begin{abstract}
We consider a dynamic network of individuals that may hold one of two different opinions in a two-party society.
As a dynamical model, agents can endlessly create and delete links to satisfy a preferred 
degree, and the network is shaped by \emph{homophily}, a form of social interaction. 
Characterized by the parameter $J \in [-1,1]$, the latter plays a role similar to Ising spins: 
agents create links to others of the same opinion with probability $(1+J)/2$, and delete them with probability $(1-J)/2$. 
Using Monte Carlo simulations and mean-field theory, we focus on the network structure in the steady state. We study the effects of $J$ on degree distributions and 
the fraction of cross-party links.
While the extreme cases of homophily or heterophily ($J= \pm 1$) 
are easily understood to result in complete polarization or anti-polarization, 
intermediate values of $J$ lead to interesting features of the network. Our model exhibits the intriguing feature of an ``overwhelming transition'' occurring when communities of different sizes are subject to sufficient heterophily: agents of the minority group are oversubscribed and their average degree greatly exceeds that of the majority group. In addition, we introduce an original measure of polarization which displays distinct advantages over the 
commonly used average edge homogeneity.
\end{abstract}

\maketitle
\section{introduction}
\begin{figure*}
     \centering
         \includegraphics[width=\textwidth]{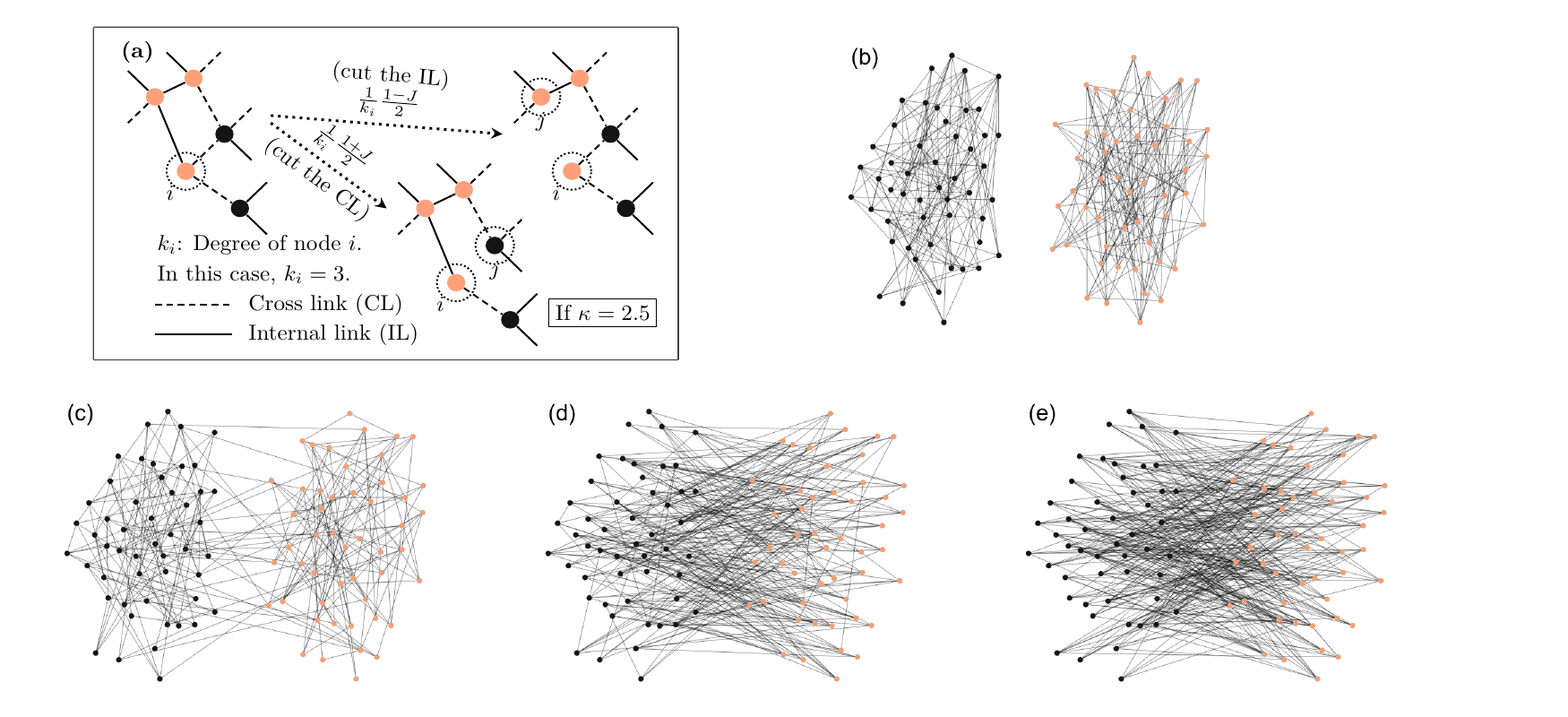}    
        \caption{(a) Illustration of the link update rule with preferred degree $\kappa=2.5$ and homophily parameter $J$. Here,  a
        node $i$ of degree $k_i=3> \kappa$ first selects one of its neighbors uniformly (node $j$); in the next time step, the link $ij$ is cut with probability $(1- J)/2$ if $\sigma_i=\sigma_j$ [$ij$ is an internal link (IL)], and with probability $(1+ J)/2$ if $\sigma_i=-\sigma_j$ [$ij$ is a cross link (CL)]. CLs and ILs 
        are necessarily cut only when $J=1$ (extreme homophily) and $J=-1$ (extreme heterophily), respectively. Similarly, if  
        $k_i<\kappa$ (not shown here), 
        then in the next time step an IL
        $ij$ is  created with probability $(1+ J)/2$ with a new neighbor $j$ of the same opinion ($\sigma_i=\sigma_j$), while a CL $ij$ is  created with probability $(1- J)/2$ with a new dissenting neighbor ($\sigma_i\neq \sigma_j$).
        (b)-(e) Different polarization scenarios after $100$ Monte Carlo steps (MCS, 1 MCS = $N$ update steps) starting with an empty graph (i.e., no links). Here,
        light and dark dots are voters (or nodes) $i$ and $i'$ holding opinion $-1$ ($\sigma_i=-1$) and $+1$ ($\sigma_{i'}=+1$), respectively. Here, 
        $N=100$,  $m=0$ (communities of same size: $N_+=N_-=50$), $\kappa=4.5$ in panels (b)-(e), and (b) $J=1$, (c) $J=0.5$, (d)  $J=-0.5$, (e) $J=-1$; see text.}
        \label{fig:Fig1}
\end{figure*}
Simple individual-based models have been commonly used to describe emergent social phenomena~\cite{Schelling-book}. 
Statistical physics models have proven particularly useful to characterize collective
behaviors of interacting populations~\cite{Castellano-rev,Galam-book,Sen-book,SW-2019}.
In the past two decades, there have been numerous advances in understanding 
the properties of these interdisciplinary models notably on social networks~\cite{Albert-rev,Dorogovtsev-book,newman2018networks}. 
An important line of research has focused on dynamical processes on networks, particularly on opinion dynamics~\cite{Castellano-rev} and evolutionary processes~\cite{Szabo-rev}.  
In this context, the dynamics of paradigmatic statistical physics models have been studied on complex networks
whose structure is random but static; see, e.g., Refs.~\cite{Antal-2006,sood2008voter,Baxter-2008,Blythe-2010,Castellano-2010,moretti2013mean,Szolnoki-2014,Sabsovich-2017}. 
In other models, collective phenomena emerge from the interactions between agents whose links 
evolve while the states of the nodes (e.g., representing agents' ``opinions'') remain static. This is for instance the case when
 individuals are more likely to bond and create links as they are more like-minded; see, e.g., Refs.~\cite{Newman-2003,Boguna-2004,Murase-2019}, 
 a form of social interaction referred to as ``homophily'' or ``assortative mixing'' ~\cite{mcpherson2001birds,centola2011experimental,centola2007complex,del2017modeling,centola2007homophily,newman2018networks}. 

Social networks are comprised of individuals with a variety of attributes, such as opinion, race, age, educational background, or gender~\cite{volkovich2014,zeltzer2020gender,mcpherson1987homophily}.
The level of homophily in a society thus reflects the tendency of individuals to establish ties with those 
having similar attributes to theirs rather than with others~\cite{mcpherson2001birds,yavacs2014impact,shalizi2011homophily,zeltzer2020gender}. This phenomenon, reminiscent of filter bubbles and echo chambers, is 
commonly seen in political parties~\cite{pariser2011filter,Iyengar2012,barbera2015tweeting,barbera2015,bakshy2015exposure,del2016spreading,wang2020public}.
Similarly, {\it heterophily} refers to 
the tendency to establish 
links  between agents with different  attributes (or dissenting ``opinions'')~\cite{xie2016,ramazi2018,barranco2019,yokomatsu2021}.
Studying how homophily and/or heterophily influence the network structure has gained importance in both sociological~\cite{yavacs2014impact,shalizi2011homophily,mcpherson1987homophily,zeltzer2020gender,yavacs2014impact,gargiulo2017role,centola2007complex,centola2011experimental,centola2007homophily} and physics-oriented literature~\cite{Boguna-2004,wong2006spatial,karimi2018homophily,kimura2008coevolutionary,papadopoulos2012popularity,asikainen2020cumulative,krapivsky2021divergence}.
In this context, homophily often features in so-called nodal attribute models~\cite{Boguna-2004,wong2006spatial,gorski2020} and growing networks like those of Refs.~\cite{papadopoulos2012popularity,karimi2018homophily,gargiulo2017role,overgoor2019choosing}, where it is generally modeled by means of a 
biased probability of adding a link or by rewiring an edge.
Homophilic interactions are often considered together with the process of structural balance~\cite{heider58}, which aims at  
eliminating tensions between a set of three connected agents (triad) by the principle of ``triad closure''~\cite{gorski2020,asikainen2020cumulative}. By combining homophilic edge weighting
and triad closure, it was recently shown that a transition to a state 
of global cooperation can occur~\cite{gorski2020}.
In Ref.~\cite{asikainen2020cumulative}, it is shown that  homophilic rewiring combined with triad closure leads to ``homophily amplification'', a phenomenon in which agents within a same group are likely to interact and establish further connections.
Furthermore, voter-like models  evolving on a network whose
links are dynamically updated according to a homophilic rewiring 
process~\cite{holme2006nonequilibrium,evans2007exact,vazquez2008analytical,vazquez2008generic,lindquist2009network,durrett2012graph,henry2011emergence}
are characterized by a continuous phase transition yielding the
fragmentation~\cite{vazquez2008analytical,vazquez2008generic}, or fission~\cite{durrett2012graph}, of the network into disconnected groups holding the same opinion.

Here, we consider an evolving network model in which links 
fluctuate continuously as the result of the homophilic or heterophilic interactions between
individuals of two communities (e.g., political parties). Contrary to most previous works on networks with homophily, the dynamics shaped by homophily here follows an evolutionary process characterized by the continuous creation and deletion of edges, with an endlessly fluctuating number of links. 
More specifically, we adopt the language of opinion dynamics and 
consider an individual-based network model 
where agents hold one of two different  opinions~\cite{mobilia2015nonlinear,castellano2009nonlinear,mellor2017heterogeneous,mobilia2007role}, and form dynamical links to satisfy a prescribed preferred degree~\cite{liu2013modeling,liu2014modeling,bassler2015networks}.  
The model dynamics can therefore be thought of as a  ``birth-death process'' for links, with transition rates depending on a homophily parameter
characterizing the interactions between nodes. 
As other preferred degree networks
(PDNs)~\cite{liu2013modeling,liu2014modeling,bassler2015networks}, our model is characterized by a nontrivial out-of-equilibrium stationary state.
By combining analytical means and simulations, 
we determine how the homophily shapes the  long-time network structure, typically characterized by the degree distributions and the fraction of cross-party links. We also quantify the extent of division between the communities by computing the network's polarization. This allows us to show that our model shares  some features found in earlier works, such as 
a fragmentation (or fission) transition under extreme homophily [see Fig.~\ref{fig:Fig1}(a) and below]. More importantly, we also show that our model exhibits intriguing features
such as an ``overwhelming transition'' occurring
when communities of different sizes are subject
to sufficient heterophily: agents of the minority group are oversubscribed and their average degree greatly exceeds that of the majority group. 

The plan  of the paper is as follows: the general formulation of the model based on  PDN dynamics with homophilic interactions
is introduced in the next section. In Sec.~ III, by combining a mean-field analysis and Monte Carlo simulations, we present  a thorough study of the model's properties when both parties are of the same size: 
the fractions of  cross-party links and of agents adding links is obtained in  Sec.~III(a), while Sec.~III(b) is dedicated to the network's degree distributions. In Sec.~IV, we consider the general case of communities of different sizes: in Sec.~IV(a), we show that  under sufficient heterophily the network consists only of agents deleting nodes; while the model's polarization is discussed in
 Sec.~IV(b). In Sec. V, we introduce a quantity that efficiently measures the network's polarization.
The final section is dedicated to a discussion of our results and to our conclusions.

%
%

\section{model formulation and general properties}
Our model is an undirected dynamical network consisting of $N$ nodes (or agents/voters) that are of two types: a fraction~$n_+$ of them is in state~$+1$, while the remaining fraction $n_-=1-n_+$ is in
state~$-1$. Hence, the population consists of number $N_{\pm}=N n_{\pm}$ agents holding opinion $\pm 1$. In the language of opinion dynamics, each node~$i$ is a ``voter'' 
whose opinion is the binary random variable $\sigma_i\in \{-1,+1\}$, i.e., each voter belongs to either party $-1$ or $+1$.  For simplicity, here $\{\sigma\}$ are quenched variables, i.e., voters are 
``zealots''~\cite{mobilia2007role,Mobilia-2013,mobilia2015nonlinear,mellor2017heterogeneous} 
(see also Refs.~\cite{Mobilia-2003,Galam-2007,SW-2011,Acemoglu-2013}). The average opinion, often referred to as ``magnetization,''
across the network is $m=\frac{1}{N}\sum_{i=1}^{N} \sigma_i=n_+ -n_-$, so that $n_{\pm}=\frac{1}{2}(1\pm m)$. 
Hence, when the magnetization vanishes, $m=0$,
each party is a group of the same size, that is $N_+=N_-$.

According to the PDN dynamics,
every node is assigned a preferred degree~$\kappa$, 
a value each agent attempts to achieve by cutting or adding links~\cite{liu2012extraordinary}. The update rules of the model, illustrated 
 in Fig.~\ref{fig:Fig1}(a), are thus as follows: at each update step, an agent~$i$ of degree $k_i$ is chosen randomly and
\begin{compactitem}
\item if $k_i>\kappa$, then the node~$i$ chooses a neighbor $j$ with
uniform probability among all its neighbors, and then either (i) the $ij$ link is
cut with the probability $\frac{1}{2}(1-J\sigma_i\sigma_j)$, or (ii) the $ij$ link remains unchanged.
\item if $k_i<\kappa,$ then the node~$i$ chooses uniformly a random node $j$ to which it is not already connected, and then either (i) the new link $ij$ is added with probability $\frac{1}{2}(1+J\sigma_i\sigma_j)$, or (ii) $i$ and $j$ remain unconnected. 
\end{compactitem}

Nodes with degrees greater than or less than~$\kappa$ are referred to as {\it cutters} and {\it adders}, respectively. 
We always take $\kappa$ to be a {\it half integer}, with 
$1 \ll \kappa \ll N$~\cite{liu2013modeling,bassler2015networks}.
This guarantees that the network is always 
dynamic, with an endlessly fluctuating number of links, and  each agent's neighborhood is a small subset of the population.
Here, $J$~is our homophily control parameter, with $-1\leq J\leq 1$.

A distinctive feature of this PDN with homophily is its ``evolutionary
dynamics'' shaped by homophily: links are continuously created and removed, 
as in a birth-death process, with rates 
capturing
the homophilic ($J>0$) or heterophilic ($J<0$) agent interactions, see Eqs.~(\ref{eqn:degmaster}) and~(\ref{eqn:R}).
%
%
As illustrated in Fig.~\ref{fig:Fig1}(a), the probability of cutting a link
between two nodes is $(1-J)/2$ if their opinions are the
same, and $(1+J)/2$ if their opinions are opposite.
It is therefore clear that $J>0$ models homophily, as it
favors the addition of {\it internal links} (ILs) between similar nodes, and the removal 
of {\it cross links} (CLs) between nodes of different opinions.
Similarly, having $J < 0$ represents heterophily that favors the creation of CLs over ILs. 

While we focus on intermediate homophily,
$-1<J<1$, with links being continuously added and cut 
 in an endlessly fluctuating network, 
a system with extreme homophily or heterophily ($J = \pm 1$) is interesting as these are the only values of $J$ for which the addition or deletion of links occurs with probability one.
In fact, 
nodes only add ILs if $J = 1$ (CLs if $J = -1$), and the network
settles in a nontrivial static configuration when every node has $k > \kappa$ and no adders are left. 

This simple model is out of thermal equilibrium, as it
violates detailed balance, and its stationary properties are thus expected to be nontrivial~\cite{liu2013modeling}, as illustrated by Fig.~\ref{fig:Fig1}(b)-\ref{fig:Fig1}(e).  
Our goal is to understand how homophily shapes the properties of the steady-state network by focusing on the total degree distributions, the fractions of CLs and adders in the stationary state of the network.
The total numbers of CLs and ILs are denoted by $L_\times$ and $L_\odot$ respectively. We can also write $L_\times=L_{+-}=L_{-+}$ and $L_{\odot}=L_{++}+L_{--}$, where $L_{\sigma\sigma'}$ is the number of links between communities holding opinion $\sigma$ and $\sigma'$. These quantities are time-fluctuating variables and the 
total number of links in the network denoted by $L=L_\times+L_\odot$
is {\it not}  conserved~\cite{liu2013modeling,liu2014modeling}. The fraction of CLs in the network is defined as $\rho=L_\times/(L_\times+L_\odot)$. When groups of $\pm 1$ voters are of different sizes ($m\neq 0$), we shall see that it is useful to distinguish 
in each community
the fraction of nodes that have CLs, see Secs.~IV and V.

We can gain some insight into the effect of~$J$ on the network dynamics by considering
the special cases
$J=\pm 1$ and $J=0$. When $J=1$, a node adds only ILs and cuts only CLs, which leads to the
population being split into two separate groups, see Fig.~\ref{fig:Fig1}(b). This phenomenon,
sometimes termed ``fission'' or ``fragmentation", where there are no CLs ($L_\times=0$, $\rho=0$),
is found in models with rewiring~\cite{vazquez2008generic,durrett2012graph,holme2006nonequilibrium,henry2011emergence}, 
and corresponds {\it complete polarization} of the population. When $J=-1$, each node can only 
add CLs and cut ILs, which eventually results in {\it complete antipolarization}, i.e., a bipartite graph with $L_\odot=0$, $\rho=1$, as in  Fig.~\ref{fig:Fig1}(e).
When $J=0$, a node adds and cuts links randomly, regardless of its neighbor's opinion, leading to a state of {\it no polarization}, where on average half of the nodes are adders, and
the average ratio of CLs to ILs, controlled purely by phase space, is $\frac{2n_{+}n_{-}}{1-2n_{+}n_{-}}$. When $0<J<1$,
the two communities are partly divided, with a majority of ILs ($\rho<1/2$); see Fig.~\ref{fig:Fig1}(c).
Similarly, when $-1<J<0$,
the network consists of a majority of CLs ($\rho>1/2$); see Fig.~\ref{fig:Fig1}(d). Hence,  the two communities are partly divided when $-1<J<1$, which results in a partial polarization of the network.

A common measure of polarization, sometimes referred to as 
``average edge homogeneity''~\cite{prasetya2020model,del2016spreading}, is the difference between the fraction of ILs and CLs, here denoted by
$\Lambda \equiv (L_\odot-L_\times)/L=1-2\rho \in [-1,1]$. When $m=0$, $\Lambda$ follows
homophily closely with  $\Lambda= 0, \pm 1$ when  $J=0, \pm 1$, respectively.
However, as shown below, its suitability deteriorates when $m$ deviates from zero: we find that for $m\ne 0$, the network can be ``polarized'' ($\Lambda>0$) even when $J=0$. In Sec.~V, we will hence  introduce an alternative measure of polarization here denoted by $\Pi$.

\section{Symmetric case, \texorpdfstring{$m=0$}{m=0}}
\begin{figure}[tbp]
\centering
\includegraphics[width=0.48\textwidth]{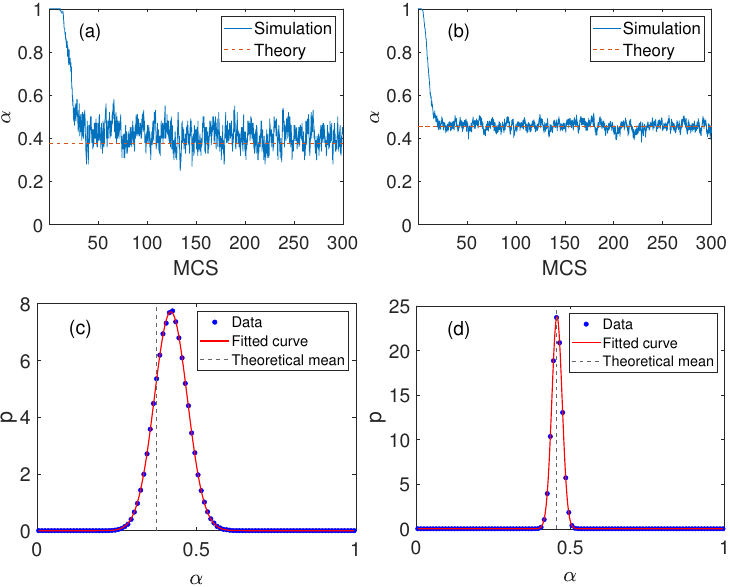}   
\caption{Evolution and probability density  of
$\protect\alpha$, the fraction of adders,  when $m=0$ for different values of $N,~\kappa$ and $J$: (a, b)
typical sample paths of  $\alpha$ as function of the time measured in the number of Monte Carlo steps (1 MCS $=N$ update steps); (c, d) stationary probability density $p(\alpha)$.
Parameters are
$(N,~\kappa,~J)=(100,~20.5,~0.5)$ in panels (a, c) and  
$(N,~\kappa,~J)=(1000,~10.5,~0.3)$ in panels (b, d).
Dashed
lines show the mean-field prediction $\protect\alpha =(1-J^{2})/2$; see Eq.~(\ref{eqn:rho}). 
Solid lines in panels (b, d) are the fitted Gaussians referred to in the text. Simulation data in panels (c, d) are obtained between 100 MCS and 5000 MCS, sampled at the end of update steps of each MCS. Data are sampled similarly in the following figures.
}
  \label{fig:Fig2}
\end{figure}

In this section, we focus on the symmetric case $m=0$
where the communities of agents holding opinion $\pm 1$ are of the same size. 
This symmetry greatly simplifies the analysis: the statistical properties of 
both communities are the same, and there is no need to distinguish between the opinion groups.  
Using mean-field analysis and Monte Carlo simulations, 
we obtain a detailed characterization of the effect of homophily on the fractions of adders and cross-links in the networks, and on 
the total and joint degree distributions.

\subsection{Fractions of adders and cross-links}
The network being dynamic, the fraction of adders, here denoted by $\alpha$, endlessly fluctuates. To gain some insight into its distribution, it is useful to start by considering the results of some typical 
Monte Carlo simulations; see Fig.~\ref{fig:Fig2}.
In all our simulations, without loss of generality,  we assume that 
there are initially no links
(e.g., mimicking a population of arriving university
students establishing links) and the number of nodes $N$ is even. 
When $m=0$, the network thus consists of 
$N/2$ nodes of each opinion. Using a range of $J$, $\kappa$, and $N$ (with $\kappa \ll N$),
we let the system evolve and perform a large number of update steps, $N$
of which correspond to one Monte Carlo step (MCS), so that in one MCS each node is picked once on average for an update move. We have noticed that after typically $\mathcal{O}(\kappa)$ MCS, the quantity $\alpha$ (as well as other global quantities such as $\rho$) reaches 
a well-defined steady state 
in which the amplitude of the fluctuations decreases as the number of nodes $N$ increases, see Figs.~\ref{fig:Fig2}(a) and~\ref{fig:Fig2}(b). In fact, $p(\alpha)$, the stationary probability density of $\alpha$,
is well fitted by a Gaussian, as shown in Figs.~\ref{fig:Fig2}(c) and~\ref{fig:Fig2}(d),  
where $p(\alpha)=7.887\mathrm{exp}[{-(\frac{\alpha-0.4208}{0.07143})^2}]$
when $N=100$ and the density is narrower when $N=1000$, in which case 
$p(\alpha)=23.97\mathrm{exp}[{-(\frac{\alpha-0.4565}{0.02348})^2}]$.
We have obtained similar results for $\rho$: the fraction of CLs also attains stationarity after 
 $\mathcal{O}(\kappa)$ MCS.
Throughout, with $\kappa$'s ranging from $4.5$ to $70.5$, we have run simulations for at least ten times longer than the time ${\cal O}(\kappa)$ necessary to reach stationarity. We then  
 collected data in the steady state 
for various lengths of time. 
In our simulations, we have thus found that
$\alpha$ and $\rho$ approach a steady state in which 
samples separated by one MCS are essentially uncorrelated, where fluctuations scale as $1/\sqrt{N}$, and the total number of links $L$ is of order ${\cal O}\left( N\kappa\right)$. While $L$, $L_\times$, $L_\odot$ are time-dependent quantities, they fluctuate around their stationary values  $\langle L \rangle$, $\langle L_\times\rangle , \langle  L_\odot\rangle$. In what follows, for notational simplicity, $L$, $L_\times$, $L_\odot$ quantities will refer to  their stationary values. Similarly, in Sec. V,  $L_{\pm \pm}$ denotes $\langle L_{\pm \pm}\rangle$.
\begin{figure}
     \centering
         \includegraphics[width=0.3\textwidth]{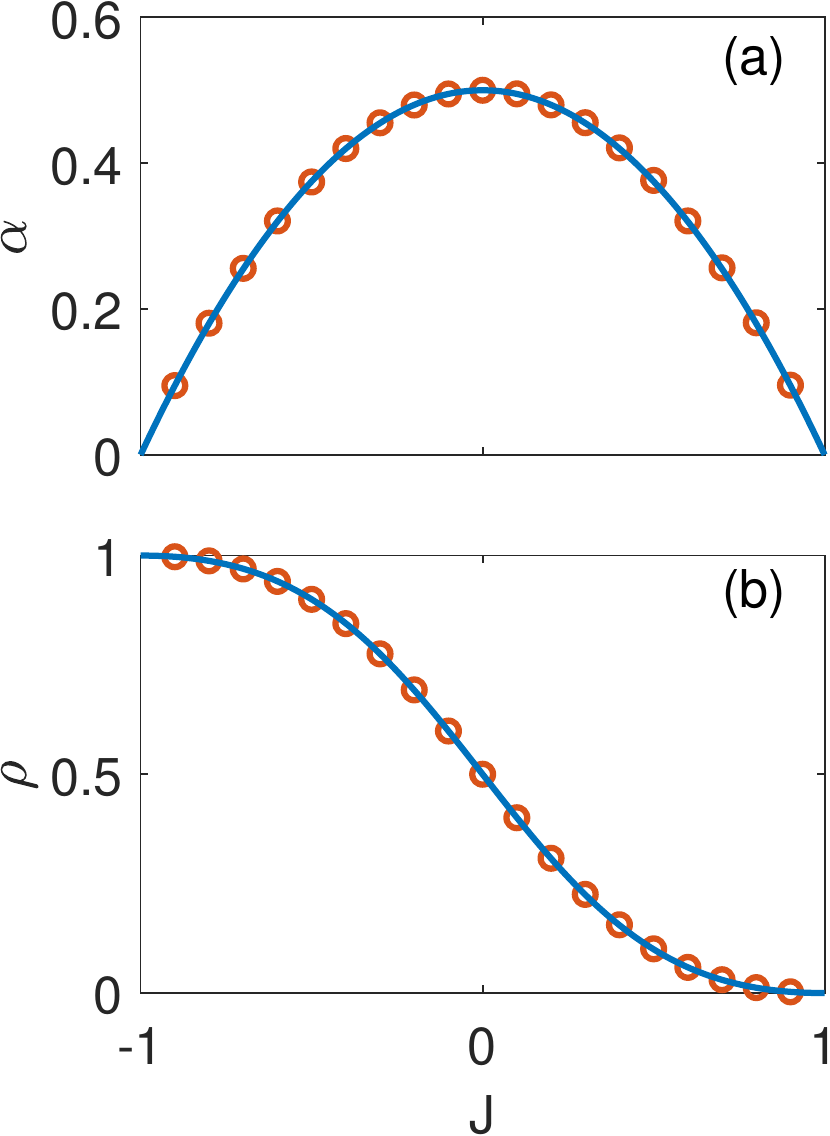}    
        \caption{Fractions of adders $\alpha$
        and of CLs $\rho$ in the steady state 
        as functions of the homophily parameter for
         $N=1000$, $\kappa=4.5$, and $m=0$. (a)  $\alpha$ vs. $J$: solid line
        is from Eq.~(\ref{eqn:alpha}), symbols are obtained by averaging simulation data collected between 100 MCS and 5000 MCS.  
        (b) Same for $\rho$ vs. $J$: solid line
        is from Eq.~(\ref{eqn:rho}). 
     }
        \label{fig:Fig3}
\end{figure}

The simulation results of Fig.~\ref{fig:Fig2} strongly
support a mean-field analysis in which the fraction of adders and 
CLs would simply be described by their stationary average values. 
When $-1<J<1$, we obtain mean-field predictions for these values, simply referred to as  $\alpha$ and $\rho$, by balancing  
the tendency for $L_\times$ and $L_\odot$ to increase and decrease in the stationary state.
For $L_\times$ to increase, an adder, picked with probability $\alpha$, must interact with a nonneighbor of different 
opinion with probability $1/2$ (since $n_{\pm}=1/2$), adding a link
with probability $(1-J)/2$. Similarly, for  $L_\times$ to decrease,
a cutter, picked with probability  $1-\alpha$, must interact with
one of its dissenting neighbors (with probability $\rho$), cutting the link with probability $(1+J)/2$. Balancing these contributions leads to $\alpha (1-J) = 2 (1-\alpha)(1+J)\rho$. Similar considerations for 
the changes in $L_\odot$ lead to 
the following additional equation
$\alpha (1+J) = 2 (1-\alpha)(1-J)(1-\rho)$. 
Solving the balance  equations for $L_\times$ and $L_\odot$, yields
 the mean-field predictions 
\begin{eqnarray}
 \label{eqn:alpha}
 \alpha&=&\frac{1}{2}(1-J^2),\\
\label{eqn:rho}
 \rho&=&\frac{1}{2}-\frac{J}{1+J^2}.
 \end{eqnarray}
Results reported in  Figs.~\ref{fig:Fig2} and \ref{fig:Fig3} show that  when $1\ll \kappa\ll N$,
the mean-field predictions for $\alpha$ and $\rho$
are in excellent agreement with 
values obtained by averaging over 
simulation data, for all values of $J$.
It is worth noting the consistency of Eqs.~(\ref{eqn:alpha}) and (\ref{eqn:rho}) 
with the consideration of the special cases above: $\rho$ increases from $0$ (complete polarization) to~$1$ (complete
antipolarization) as the homophily parameter~$J$ decreases from $1$ to~$-1$. At the two extremes, $J=\pm 1$,
the fraction of adders~$\alpha$ is zero. In the absence of homophily, $J=0$, the fractions of
CLs and adders are~$1/2$, and there is no polarization.

\subsection{Total and joint degree distributions}

\begin{figure}
     \centering
         \includegraphics[width=0.35\textwidth]{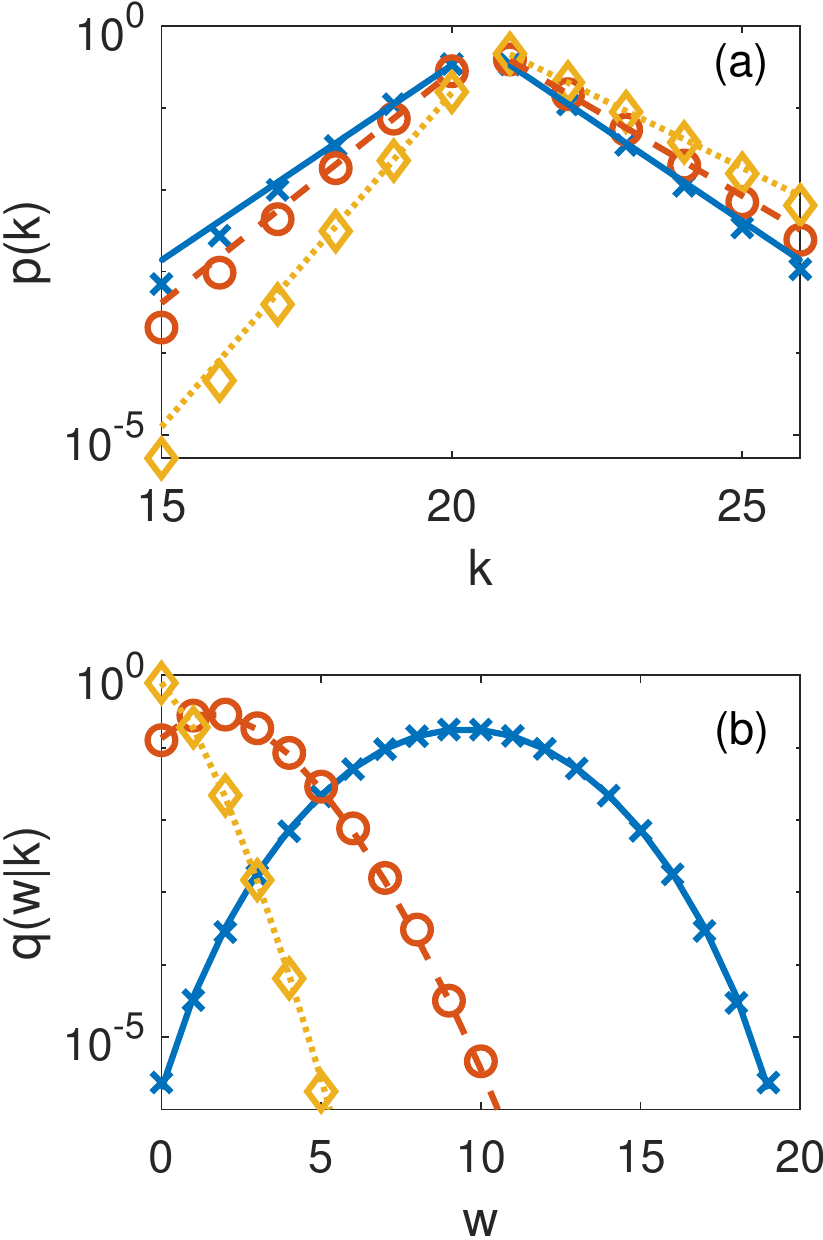}    
        \caption{  (a) Total degree distribution $p(k)$ vs. degree $k$ for $N=1000$, $\kappa=20.5$, $m=0$, and different values of $J$.
        Solid, dashed, and dotted lines are from Eq.~(\ref{eqn:finalpk}) for different values of $J$.
        (b)  Conditional distribution $q(w\,|\, k)$ vs. degree $w\leq k$ for
     $k=19$ and different values of $J$. Lines are predictions from the binomial distribution Eq.~(\ref{eqn:binomial}). Symbols represent data collected between 2000 MCS and 1000000 MCS. In both panels, $J=0$ ($\times$, solid lines), $0.5$ ($\circ$, dashed lines), $0.8$ ($\diamond$, dotted lines). 
     }
        \label{fig:Fig4}
\end{figure}

In addition to $\alpha$ and $\rho$, we are interested in determining the effect of $J$ on the long-time network structure, characterized by its degree distributions. In this section, we investigate the total degree distribution (giving the probability for a node to have degree $k$ in the stationary state), shown in Fig.~\ref{fig:Fig4}(a), and the conditional degree distribution, shown in Fig.~\ref{fig:Fig4}(b). The former can be obtained by combining the above mean-field
theory with a master-like equation obeyed by
the degree distribution
at time~$t$, denoted by $p(k,t)$. If $R^a(k)$ and $R^c(k)$ are the rates at which a node of degree~$k$  adds or cuts a link, then
$p(k,t)$~obeys
\begin{align}
\label{eqn:degmaster}
  \frac{d p\left(k,t\right)}{dt} =&R^{a}\left( k-1\right) p\left(k-1,t\right) +R^{c}\left( k+1\right)p(k+1,t)\nonumber\\
 &-\left[ R^{a}\left( k\right)
 +R^{c}\left( k\right) \right] p\left( k,t\right).
 \end{align}
Since this master equation governs a single-variate distribution, the steady state $\lim_{t\to \infty}p(k,t)=p(k)$ is obtained by balancing the probability that a node of degree $k$ acquiring a link 
[with rate $R^a(k)$] with that of the node of degree $k+1$ losing a link [with rate $R^c(k+1)$], i.e.,  
$R^{a}\left( k\right) p\left( k\right) =R^{c}\left( k+1\right) p\left(k+1\right)$.
Once we determine expressions for $R^a$ and  $R^c$, the recursion relation and normalization condition $\Sigma_{k} p(k) = 1$ readily gives an explicit expression for $p(k)$. 

Under PDN dynamics~\cite{liu2013modeling,liu2014modeling,bassler2015networks}, links are added and cut from a node both actively (action by the chosen node)
or passively (action by other agents). Specifically, when $k>\kappa$, a node increases its degree only passively. 
In one time step, an adder can be chosen with probability $\alpha$ and a link added 
with probability $\frac{1}{2}=\frac{1}{2}[\frac{1}{2}(1+J)+\frac{1}{2}(1-J)]$ (assuming that half of the 
nonneighbors are of the same (or different) opinion when $\kappa\ll N$). Hence, in the spirit of the mean-field approximation, this yields $R^{a}= \frac{\alpha }{2}$. For $R^{c}$, similar reasoning
leads to the probability for cutting a link being $\chi (\rho )\equiv\frac{1}{2}%
(1-J)(1-\rho )+\frac{1}{2}(1+J)\rho $. Since $k>\kappa $, a node can take
this action, as well as suffer a decrease passively, from the fraction $%
1-\alpha $ of other cutters. Thus, $R^{c}= \chi(\rho) \lbrack 1+(1-\alpha )]$%
. Similar arguments can be used when $k<\kappa $, leading to $R^{a}=
\frac{1+\alpha }{2}$ and $R^{c}= \chi(\rho) (1-\alpha )$. 
Combining these consideration, we have
 \begin{equation}
 \label{eqn:R}
 R^{a} = \frac{1}{2} [ H (\kappa - k ) + \alpha ],
 \quad
 R^{c} = \chi(\rho)[ H ( k - \kappa ) + (1-\alpha) ]
 \end{equation}
where $H$ is the Heaviside step function.
Using the mean-field results Eq.~(\ref{eqn:rho}) for $\rho$ to rewrite $\chi$
as a function of $J$ in Eq.~(\ref{eqn:R}) and solving the recursion relation, we obtain the stationary total
degree distribution as the steady-state solution of Eq.~(\ref{eqn:degmaster}):
\begin{align}
p(k)=\begin{dcases}\label{eqn:finalpk}
\left(\frac{1-J^2}{3-J^2}\right)^{\ceil{\kappa}-k}\ \text{for}\ k<\kappa,\\
\left(\frac{1+J^2}{3+J^2}\right)^{k-\floor{\kappa}}\ \text{for}\ k>\kappa.
\end{dcases}
\end{align}

Interestingly, $p(k)$ is an even function of $J$ for all $k$'s:
when $m=0$, homophily and heterophily have the same effect on the distribution of degrees in the 
stationary network. This is no longer the case when $m\neq 0$; see below. 
We notice that Eq.~(\ref{eqn:finalpk}), in accord with simulation results, predicts that $p(k)$ is symmetric with respect to $\kappa$  
when $J=0$ (no polarization): in this case, we recover a Laplace distribution as in Refs.~\cite{liu2013modeling,bassler2015networks}. 
However, $p(k)$ is skewed as soon as there is some degree of homophily ($J\neq 0$):
in Fig.~\ref{fig:Fig4}(a), the slopes of the left branch of $\ln{(p(k))}$ increase from $\ln 3$ 
to infinity, while those of the right branch increases from $-\ln 3$ to $-\ln 2$, as $|J|$ increases from $0$ to $1$. 
Comparison with simulation results shows that these predictions Eq.~(\ref{eqn:finalpk})
are in very good agreement with data over a broad range of values of $J$ ($-1<J<1$) and $k$ 
(Fig.~\ref{fig:Fig4}(a)). The deviations near the tails of the distribution are understandable, as 
our approximation does not account for the physical limits of $ k \in [0,N)$. 

\begin{figure}
     \centering
         \includegraphics[width=3.325in]{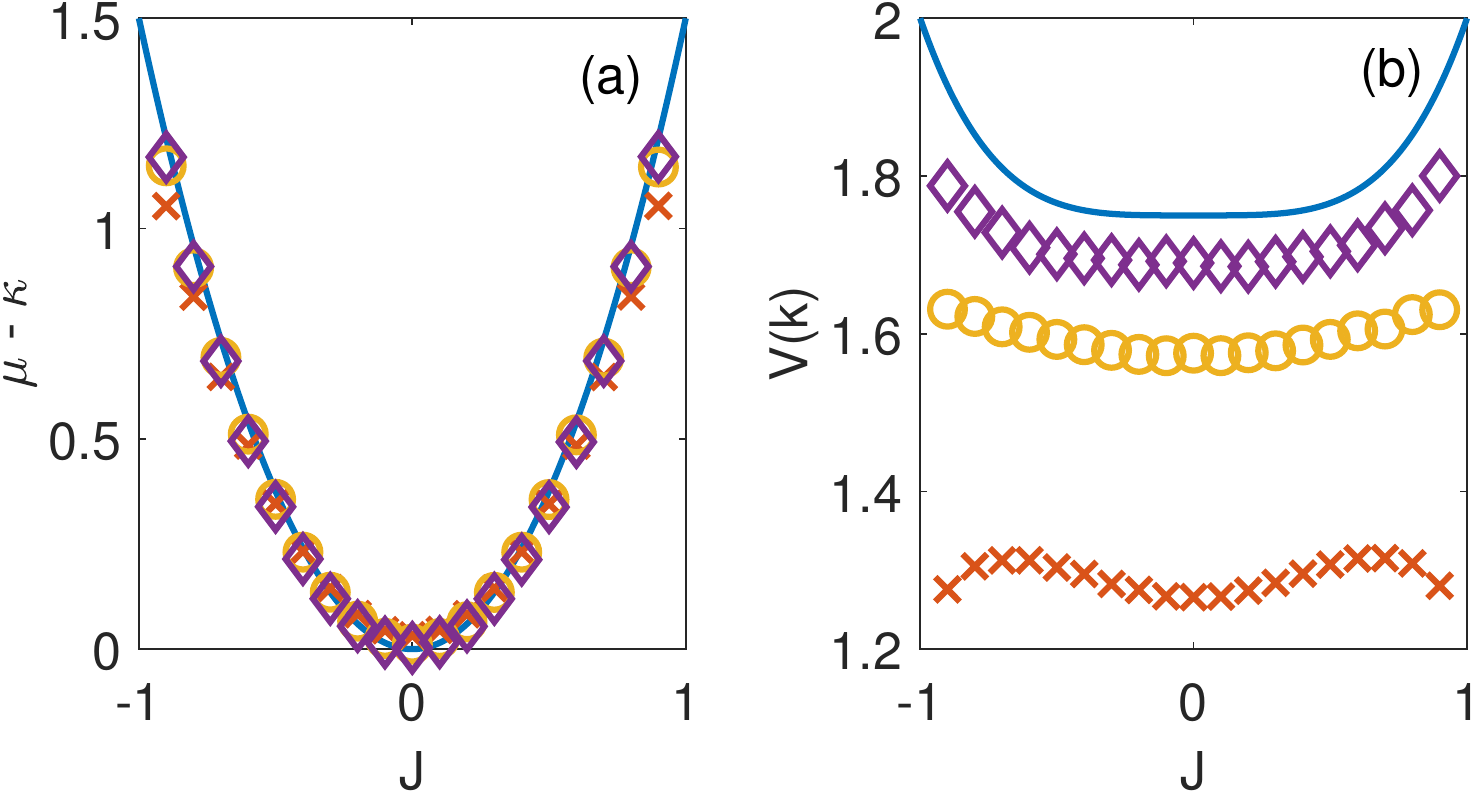}    
        \caption{(a) $\mu-\kappa$ vs. $J$ and (b) $V(k)$ vs. $J$ for $N=1000$ and different values of $\kappa$:
         $\kappa=5.5$ ($\times$), $20.5$ ($\circ$), and $70.5$ ($\diamond$). Lines are the analytical degree average and variance given by Eq.~(\ref{eqn:muV}); markers are these quantities obtained by averaging simulation data collected between 2000 MCS and 1000000 MCS, see text. 
     }
        \label{fig:Fig5}
\end{figure}

With Eq.~(\ref{eqn:finalpk}), we can compute the average degree $\mu=\sum_k k p(k)$ and variance $V =\sum_k (k-\mu)^2 p(k)$, whose  explicit expressions read 
\begin{eqnarray}
 \label{eqn:muV}
 \mu =\kappa + \frac{3J^2}{2} \quad \text{and} 
 \quad V = \frac{7+J^4}{4}.
\end{eqnarray}
The results for $\mu$ are in good agreement with those from simulation data when $1\ll \kappa\ll N$, and the results for $V$ approach the theoretical prediction for large $\kappa$, as shown in Fig.~\ref{fig:Fig5}.
We notice
that somewhat counterintuitively $\mu$ increases from $\kappa$ monotonically with $|J| $. In other words, both
homophily and heterophily increase the average degree, which is consistent with the
decrease of adders. 
More noteworthy is that the presence of translational
invariance (in $k$-space) in our approximation scheme for $p(k)$, i.e. the 
dependence on $(k, \kappa)$ is only through the difference $k-\kappa$; see Eq.~(\ref{eqn:finalpk}). As a result,  both $\mu-\kappa$ and $V$ are
independent of $\kappa$. In fact, $\mu-\kappa={\cal O}(J^2)$
and the standard deviations from the mean degree are also of order 
${\cal O}(J^2)$.
The systematic deviations from the 
theoretical prediction of $V$ in Fig.~\ref{fig:Fig5}(b) stem from
finite-size effects and decrease as $\kappa$ is set further from the limits of our  approximation ($1 \ll \kappa \ll N$). 

To summarize, our mean-field theory, resulting in Eqs.~(\ref{eqn:finalpk}) and (\ref{eqn:muV}), captures the essence of our model when $m=0$ and 
agrees well with simulation data, with some deviations caused by some of the underpinning mean-field assumptions. In particular, we have found that the total degree distribution in the case of communities of the same size
is exponential with a peak around the preferred degree $\kappa$ (when $1 \ll \kappa \ll N$) with small deviations 
about it that increase with the level of homophily or heterophily in the population. As discussed in  the next section, a totally different and more complex picture emerges when communities are of different sizes.

In addition to $p(k)$, we are also interested in how the CLs and ILs are distributed. 
We have thus studied the conditional distribution 
$q(w\,|\, k)$, giving the probability for a node of total degree $k$ to have $w$ CLs in the stationary state. As in other network models with preferred degrees~\cite{bassler2015networks}, we 
expect no bias in favor of or against a CL other than the effects of $J$, in such way to produce 
the observed value of $\rho$. 
In other words, our assumption is that, for a node with degree $k$, the probability of selecting one of its neighbors of the opposing opinion is just $\rho$. 
Hence, we may postulate a
binomial distribution for $w$, the number of CLs our node has:%
\begin{equation}\label{eqn:binomial}
q\left( w|k\right) =\binom{k}{w}\rho^{w}\left( 1-\rho \right) ^{k-w}.
\end{equation}%
The distribution is in excellent agreement with simulation data obtained for $q(w\,|\, k)$ with different sets of parameters, as illustrated in Fig.~\ref{fig:Fig4}(b). 

Note that when $J$ changes sign, $J \to -J$, we have $\rho \to 1-\rho$
and therefore  $q(w|k) \to q(k-w|k)$. These ``degeneracies''
 will be lifted once we consider communities
of different sizes. 

Beyond these comparisons, let us point out an interesting and sharp
distinction between the total degree distribution and the conditional distribution $q$. 
The variance of the former is $\mathcal{O}\left(1\right)$. Since
it is independent of extensive parameters like $N$ and $\kappa$, the total degree distribution
resembles a delta function in the large $N, \kappa$ limit. By contrast,
being a binomial in $w$, the variance of $q$ is $\rho \left(1-\rho \right)
k$. Since the $k$'s of interest are $\mathcal{O}\left( \kappa \right) $, the variance here is of extensive form,  a typical feature of random networks like Poisson random graphs~\cite{Dorogovtsev-book,newman2018networks,renyi1959random}, also found in rewiring models~\cite{vazquez2008analytical,vazquez2008generic} where the mean and the variance of degrees are of the same order.


\section{Asymmetric case, $m \neq 0$}
We now consider the general case where
opinion groups are of different sizes, with $n_{+}\neq n_{-}$, i.e., 
 $m \neq 0$. In this case, each physical quantity is twofold: the fractions of adders or cutters and the fractions of CLs or ILs have to be distinguished in each community. Similarly, the rates at which a node adds or cuts a link are different in each community.
 As a result, the general asymmetric case $m\neq 0$
turns out to be surprisingly complex,
and its thorough analysis is presented elsewhere~\cite{companion}.
Here, our main goal is to present the salient features 
of the model in this general case and to highlight an original
phenomenon, referred to as the ``overwhelming transition,'' occurring here under sufficient heterophily.
We also provide arguments explaining the original phenomenology and 
provide some insights on how to generalize the 
analysis carried out in the symmetric case $m=0$.
For this, we first discuss the fraction of adders and CLs
and then the degree distribution.
\subsection{Fractions of adders and cross-links}
\begin{figure}
     \centering
         \includegraphics[width=0.5\textwidth]{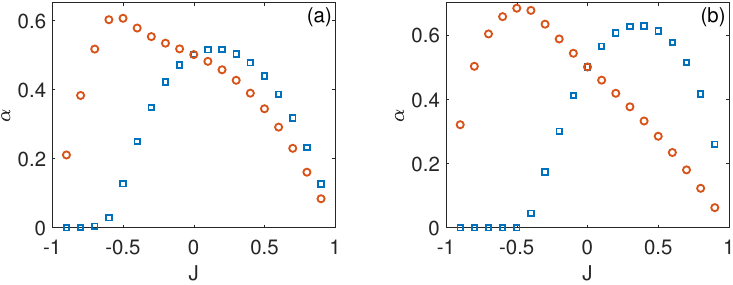}
     \caption{
     Simulation results for the fraction of adders $\alpha_+,~\alpha_-$ when $N=1000$ and $\kappa=60.5$. $\alpha_+$ (blue  $\square$) and $\alpha_-$ (red $\circ$)
     versus $J$ with (a) $m=-0.2$ and (b) $m=-0.6$. Data are collected after $10^5$ MCS.}
     \label{fig:Fig6}
\end{figure}

We denote by $\alpha_{\sigma}$ the fraction of adders 
in the communities of opinion $\sigma=\pm$. Similarly, we denote by $\rho_{\sigma}=L_{\times}/(L_{\times}+2L_{\sigma \sigma})$ the fraction of CLs in opinion group $\sigma$, giving the probability of a link connected to a node with opinion $\sigma$ being a CL. 

%
As a result of the asymmetry, in general $\alpha_+\neq \alpha_-$ and $\rho_+\neq \rho_-$, with  $\alpha_{\pm}=\alpha$ and $\rho_{\pm}=\rho$  when 
$m=0$. Hence, when $m\neq 0$, each panel of the counterpart of Fig.~\ref{fig:Fig3} contains twice as many curves, one for each community, as we see in Fig.~\ref{fig:Fig6} [to be compared with Fig.~\ref{fig:Fig3}(a)]. 
The special case $J=0$  is intuitively
simple since the system thus behaves as if there was just a single population (the 
distinction of opinion is merely nominal), so that the addition and removal of links are unbiased. Hence, $\alpha_{\pm } =1/2$
regardless of $m$. Furthermore, 
for an agent of opinion $\sigma$, the average
fraction of CLs when $J=0$ is simply the fraction of agents of the opposite opinion, yielding $\rho_{\pm} =n_{\mp}=\left( 1\mp m\right)/2$. 

Simulation results show that in general
$\alpha_{\pm}\left( m, J\right) $ and $\rho_{\pm}\left( m, J\right) $ 
are nontrivial functions of $J$ and $m$; see Fig.~\ref{fig:Fig6}, 
where we find that  $\alpha_{\pm}$ have a complex dependence on $J$
and a very different shape when $m=-0.2$ [Fig.~\ref{fig:Fig6}(a)]
and $m=-0.6$ [Fig.~\ref{fig:Fig6}(b)].

When $J>0$,
there is always a finite fraction of adders in both communities 
($\alpha_{\pm }>0$), whereas when $J<0$, the fraction of adders in the smaller group 
($\alpha_{+}$ in Fig.~\ref{fig:Fig6}) vanishes when heterophily is too
strong. In other words, when $J$ is close enough to $-1$, the minority 
{\it consists only of cutters}.
When $J>0$ and both communities are of comparable sizes ($|m|\ll 1$), we recover a scenario similar to the symmetric case,
e.g., Fig.~\ref{fig:Fig6}(a), with a fraction of cutters and adders in
both groups are comparable. However, for larger asymmetry,
the fraction of adders in the smaller community is considerably larger 
than that in the majority agents [$\alpha_{+}\gg $ $\alpha_{-}$ 
in Fig.~\ref{fig:Fig6}(b)] if $J>0$, but otherwise ($\alpha_{+}\ll $ $\alpha_{-}$)
for $J<0$. Indeed, as noted above, $\alpha_{+}$ is undetectably small when 
$J$ drops below some threshold value (for example, when $m=-0.6$, the threshold is $J\approx -0.42$). 

\subsection{Total degree distribution 
and an ``overwhelming'' transition}

\begin{figure}
     \centering
         \includegraphics[width=0.5\textwidth]{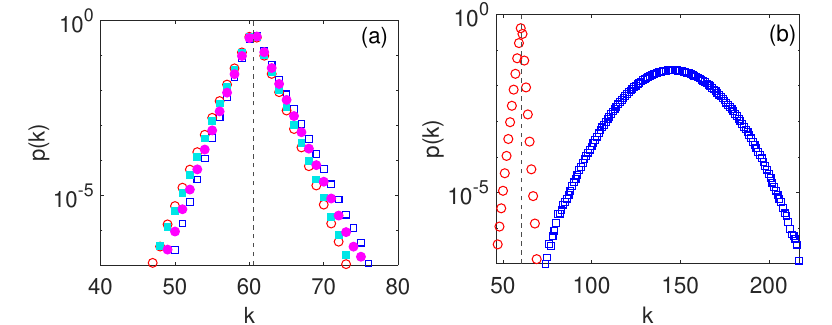}
     \caption{
     Simulation results for the total degree distributions $p_{\pm}(k)$ in the community of $\pm 1$ nodes
     for different values of $J$ and $m$ when $N=1000$ and $\kappa=60.5$. Data are collected after $10^5$ MCS. 
     Dashed lines are eyeguides showing $k=\kappa=60.5$.
     (a) Symbols $\square$ and $\circ$ refer to $p_+$ and $p_-$, respectively, for $m=-0.2$, $J=-0.2$ (blue and red empty markers) and $J=0.2$ (cyan and magenta filled markers).
     (b)  Symbols $\square$ (blue) $\circ$ (red) refer respectively to $p_{+}$ and $p_{-}$, for $m=-0.6$, $J=-0.6$, when the minority agents are ``overwhelmed'' by those in the majority (see Sec. IV). }
     \label{fig:Fig7}
\end{figure}

Turning to degree distributions, we denote by $p_{\sigma}(k)$ the
probability that an agent with opinion $\sigma$ has $k$ links in total
(regardless of the opinion of its neighbors). Figure~\ref{fig:Fig7} clearly 
illustrates that $p_{+}(k)\neq p_{-}(k)$, with nontrivial dependence of 
$p_{\sigma}(k)$
on both 
$m$ and $J$. In particular, in Fig.~\ref{fig:Fig7}(b) with strong heterophily ($J=-0.6$), we notice that 
the minority community is characterized by degrees greatly exceeding $\kappa$ and following a broad distribution.

For low asymmetry and small $|J|$, $p_{\sigma}(k)$ are
qualitatively similar to the $p\left(k\right) $ above, compare Figs.~\ref{fig:Fig4}(a) and~\ref{fig:Fig7}(a) for $m=-0.2$ and $J=\pm 0.2$. Though $p_{\pm }(k)$ are no longer even functions of $J$, they are
still (approximately) exponential distributions peaking near $\kappa$. Hence,
the mean degree of all nodes is close to the preferred $\kappa$.

By contrast, striking behavior emerges under large asymmetry and high
level of heterophily ($|m|=\mathcal{O}(1)$ and $J$ near $-1$) as illustrated
in Fig.~\ref{fig:Fig7}(b) for $m=J=-0.6$. In this case, all minority agents are cutters,
while the degree distribution  ($p_+$ in Fig.~\ref{fig:Fig7}) is {\it Gaussian-like}, {\it with a mean much
larger than $\kappa$}. However, the distribution
for a majority agent,  $p_{-}\left( k\right)$ in Fig.~\ref{fig:Fig7}, 
 is comparable to those in the cases of small $m, J$: it is approximately an
 exponential distribution peaking around~$\kappa$.
Intuitively, this intriguing behavior stems from the combined effect of the preferred degree and heterophily mechanisms resulting in the minority agents being
``overwhelmed'' by those in the majority.
In fact, 
when one group is larger than the other and strong heterophily favors the creation of CLs, agents in the smaller group can be ``overwhelmed'' by links created by members of the majority group, and their degree can exceed $\kappa$ forcing them to act as cutters. By analogy 
with the mechanism of 
``homophily amplification'' of Ref.~\cite{asikainen2020cumulative} by
which agents of the same group interact and establish further connections, this phenomenon can thus  be described as a sort of ``heterophily amplification mechanism.''

To provide a more quantitative picture, we consider $L_{\times}$, the total number of CLs. Roughly, due
to the large number of agents in the majority, these can act ``as they wish'' 
and settle with degrees around $\kappa $, which can
provide an estimate for $L_{\times}$.
If $J=0$, then $\rho _{-}=n_{+}$, so that 
$L_{\times }=\left( \kappa N_{-}\right) n_{+}$.
 However, if heterophily is
strong, then the most naive estimate of $L_{\times }$ would be larger by a factor
of $b=\left( 1-J\right) /\left( 1+J\right) $, which is the bias in
favor of making CLs. Thus, a minority node (which has opinion $+$ here) would have $L_{\times }/N_{+}\sim
\kappa n_{-}b$ CLs. Thus, for large asymmetry and heterophily, 
the number of CLs alone can greatly exceed $\kappa$.
Meanwhile, a minority agent is biased {\it against} 
cutting these CLs [suppressed by $\left( 1+J\right) /2$]. In this
scenario, minority agents are overwhelmed by the majority adding links
to them preferentially. Their cutting cannot keep up with the creation of links by the
opposing group. As a result, their degrees are significantly larger than 
$\kappa$, as seen in Fig.~\ref{fig:Fig7}(b). 
This is in striking contrast with what we have found in 
the symmetric case $m=0$, where 
the degree distribution is always centered about $\kappa$,
and shows that, when communities are of different sizes, simple update rules like those of the PDN can lead to a broad degree distribution
with a large average degree
of the smaller group. In Ref.~\cite{companion},
this picture is corroborated by a detailed analysis of the 
``overwhelming transition''
and of $p_{\pm}(k)$ in terms of suitable analytical approximations.
Interestingly, the authors of Ref.~\cite{karimi2018homophily} studied a two-community growing network according to the preferential attachment 
dynamics with homophilic interactions, and
showed that in their model heterophily helps increase the degree of the minority group, but these authors did not report the existence of an overwhelming transition in their model.


%
\section{Polarization}
\begin{figure}[!t]
     \centering
         \includegraphics[width=0.46\textwidth]{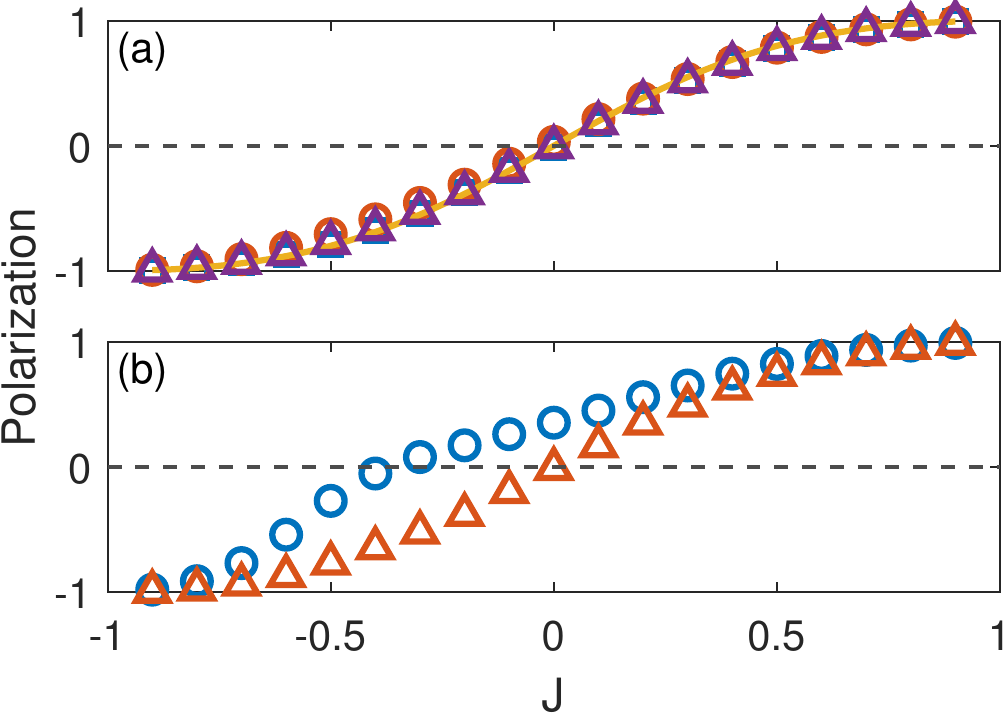}
     \caption{Measures of polarization $\Lambda$ and $\Pi$ vs. $J$ for different values of $m$. Symbols are from simulation data. (a) $\Lambda=1-2\rho$ for $m=0$ ($\square$) and $m=-0.2$ ($\circ$), and $\Pi$ for 
     $m=-0.2$ ($\triangle$). The line shows the mean-field prediction $\Lambda=2J/(1+J^2)$  obtained for $m=0$ by using Eq.~(\ref{eqn:rho}) in Eq.~(\ref{eqn:Lambda}), while $\Pi$ has been computed from
     its definition Eq.~(\ref{eqn:Pi}) using
     simulation data.  (b)  $\Lambda$ ($\circ$)  and $\Pi$ ($\triangle$) as functions of $J$ with $m=-0.6$; when $J=0$, $\Lambda=0.36$
     and $\Pi=0$; see text. The dashed line is an eyeguide showing {\it zero} polarization. 
     In all panels:  $N=100$ and $\kappa =6.5$.
     Data are collected and sampled from $10^3$ to $10^5$ MCS. }
     \label{fig:Fig8}
\end{figure}
In this section, we study 
the phenomenon of polarization that measures 
the extent of division between communities with different 
opinions. We have seen that in the case of extreme homophily ($J=1$),
there is ``fission,'' which results in  complete polarization with 
the network split into two separate communities; see Fig.~\ref{fig:Fig1}(b). Oppositely, when there is extreme heterophily ($J=-1$), the network becomes bipartite, and in this case, 
there is complete antipolarization; see Fig.~\ref{fig:Fig1}(e).
 
To characterize the level of partial 
division between the parties arising for intermediate homophily, $-1<J<1$ [Figs.~\ref{fig:Fig1}(c) and \ref{fig:Fig1}(d)], polarization is often measured in terms of the so-called average
edge homogeneity~\cite{prasetya2020model,del2016spreading}. The latter quantity, here denoted by $\Lambda$, is defined as the difference between the fraction
of ILs and CLs, that is $\Lambda=1-2\rho$. When $m=0$, it has a simple dependence on $J$ that is well captured by Eq.~(\ref{eqn:rho}), yielding $\Lambda=2J/(1+J^2)$. However, in general the 
fractions of ILs and CLs, and hence $\Lambda$, depend on the size of each group ($Nn_{\pm}$) and on $\rho_{\pm}$. In the realm of the mean-field approximation, we indeed have $1/\rho_{+}+1/\rho_{-}=2/\rho$, yielding
\begin{equation}
\label{eqn:Lambda}
\Lambda=1-2\rho =1-\frac{4\rho_{+}\rho_{-}}{\rho_{+}+\rho_{-}},
\end{equation}
%
which is a nontrivial function of $m$ and $J$; see Fig.~\ref{fig:Fig8}.
This quantity provides a meaningful measure of polarization in symmetric
communities of similar sizes, i.e. when $m$ is close to zero. In  this case, $\Lambda$ indeed captures the correct degree of polarization  $\Lambda \to \pm 1$ when $J\to \pm 1$
and $\Lambda \propto J$ when $J\approx 0$, see Fig.~\ref{fig:Fig8}(a). 
However, we note 
 that $\Lambda$ can provide misleading impressions for $m\neq 0$.
 This can be seen by noticing that when $J=0$,
  $\rho_{\pm}=(1\mp m)/2$, and Eq.~(\ref{eqn:Lambda}) thus
gives $\Lambda=m^2$, as in Fig.~\ref{fig:Fig8}(b). However, when
 $J=0$, agents not discriminating between the communities,
there is no reason to associate it with any level of polarization:
a proper measure of polarization under $J=0$ should thus give 
zero.

\begin{figure}[!t]
     \centering
         \includegraphics[width=0.5\textwidth]{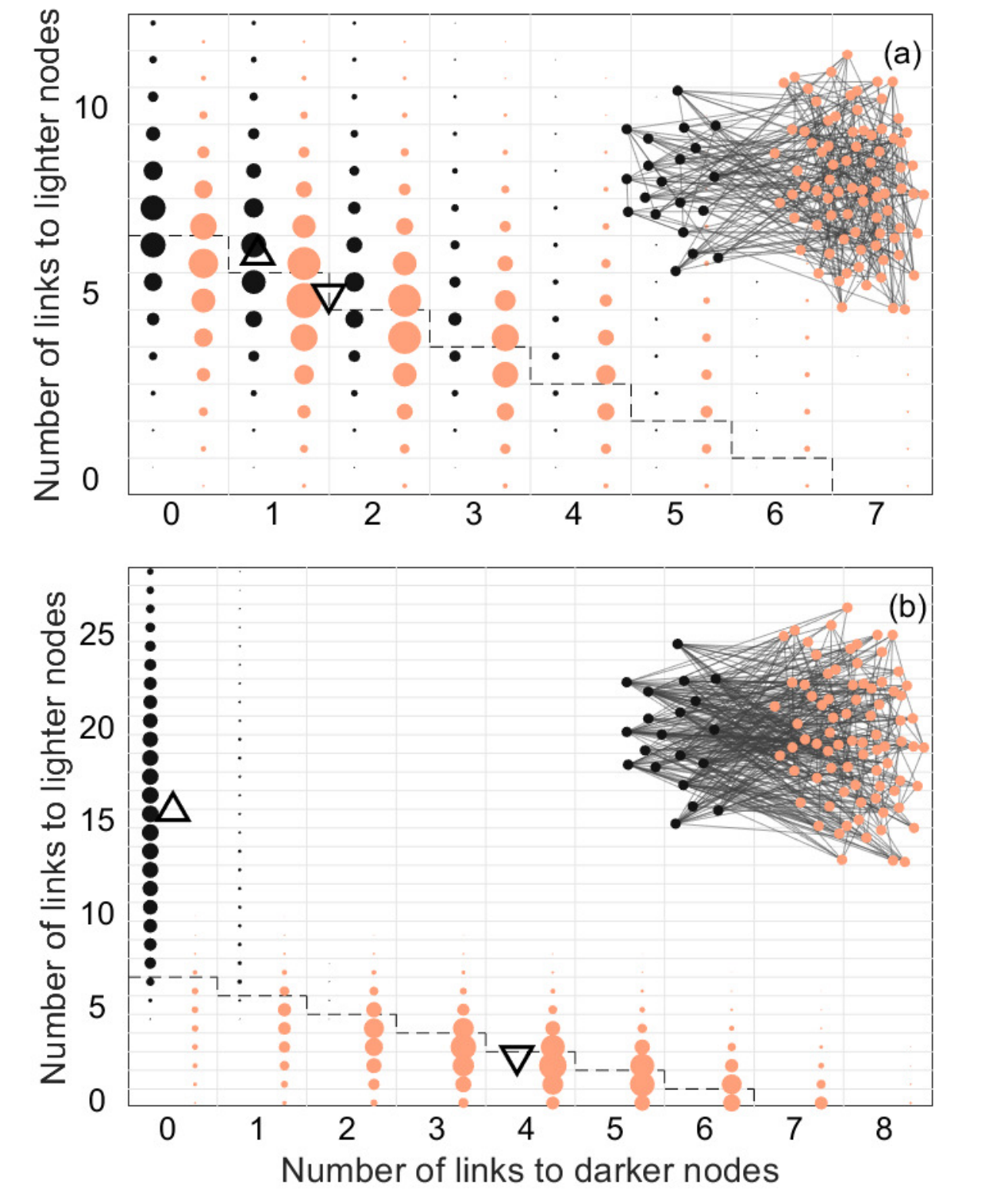}
     \caption{Visualization of the joint degree distributions $P_{+}(\ell_{+}, \ell_{-})$ (dark dots)
     and $P_{-}(\ell_{+}, \ell_{-})$ (light dots)
     with $N=100, \kappa=6.5$ and $m=-0.6$. (a) $J=-0.1$; (b) $J=-0.6$: each dot
      represent a node holding opinion $+ 1$ (dark) or $-1$ (light).
      A node located in cell
     $(\ell_{+}, \ell_{-})$ has $\ell_{+}$ and $\ell_{-}$ links to $+1$ and $-1$ nodes, respectively, and total degree $\ell_{+}+\ell_{-}=k$. The area of each dark or light 
     dot is proportional to the number of nodes having respectively $\ell_{+}$ and $\ell_{-}$ links to $+1$ and $-1$ nodes. Dashed lines show  $\ell_{+}+\ell_{-}=\kappa$.
     $\triangle$ and $\dtriangle$ show, respectively, the centers of mass $(\bar{\ell}_{+}, \bar{\ell}_{-})_{\sigma}$ of nodes with opinion $\sigma$ [e.g., $\triangle$ is the center of mass of $P_{+}(\ell_{+}, \ell_{-})$]; see text.
     Data are collected and sampled from $10^3$ to $10^5$ MCS.  Insets: illustration of typical network configurations after 1000 MCS.
          }
     \label{fig:Fig9}
\end{figure}
For a better measure of polarization, we turn to the {\it joint} degree distribution 
$P_{\sigma}(\ell _{+},\ell _{-})$. This quantity gives the probability that a node holding opinion 
$\sigma $  has $\ell _{\tau }$ links to agents with opinion $\tau=\pm $. These
distributions are illustrated in Fig.~\ref{fig:Fig9}, 
where $P_{+}(\ell _{+},\ell _{-})$ and $P_{-}(\ell _{+},\ell _{-})$
are, respectively, displayed by dark and light dots, and 
where each cell is labeled by $(\ell _{+},\ell _{-})$, with the size of the dots being proportional to $P_{\sigma }(\ell
_{+},\ell _{-})$.
The averages 
\begin{equation*}
(\bar{\ell}_{\pm })_{\sigma }\equiv \sum_{\ell _{\pm }} \ell _{\pm } P_{\sigma }
\end{equation*}%
can be regarded as the ``centers of mass'' (CMs) of 
the distributions $P_{\sigma }$. Clearly, the two CMs will not
coincide in general, as illustrated by $\triangle $ and $\dtriangle$
in Fig.~\ref{fig:Fig9} (where they have been obtained from simulation data).
Nevertheless, it can be shown that they do coincide when $J=0$, where the opinions of the nodes are irrelevant~\cite{companion}.
%
Thus, the separation between the two CMs can serve as a
suitable measure of polarization. Specifically, we define a ``normalized'' distance between CMs and measure of  polarization as 
\begin{equation}
\label{eqn:Pi}
\Pi \equiv%
\frac{1}{2}\left\{\frac{(\bar{\ell}_{+})_{+}-(\bar{\ell}_{+})_{-}}{(\bar{\ell}_{+})_{+}+(\bar{\ell}_{+})_{-}}+\frac{(\bar{\ell}_{-})_{-}-(\bar{\ell}_{-})_{+}}{(\bar{\ell}_{-})_{-}+(\bar{\ell}_{-})_{+}}\right\}.
\end{equation}%
In the case of complete polarization, there are no CLs, so
$(\bar{\ell}_{\sigma})_{-\sigma}$ vanishes and $\Pi =1$. Similarly, for antipolarization, there are no ILs, so
$(\bar{\ell}_{\sigma })_{\sigma}$ vanishes and $\Pi =-1$. Furthermore, since
$(\bar{\ell}_{\sigma})_{\sigma}=(\bar{\ell}_{\sigma})_{-\sigma}$ we have $\Pi =0$ when $J=0$ for all $m$. In other words, this definition of polarization vanishes in the absence of homophily for any asymmetry in the population sizes, and avoids the deficiencies of $\Lambda$. The quantity Eq.~(\ref{eqn:Pi}) has therefore the required properties to meaningfully characterize  
 polarization in networks with communities of arbitrary sizes.

Figure~\ref{fig:Fig8} illustrates the salient features of $\Pi$ with simulation results.
We find that when $|m|\ll 1$, both $\Lambda$ and $\Pi$ are well
approximated by $\Lambda \approx \Pi \approx 2J/(1+J^2)$, where
we have used the mean-field expression Eq.~(\ref{eqn:rho})
for $\rho$ in Eq.~(\ref{eqn:Lambda}). With $m=-0.2$, Fig.~\ref{fig:Fig8}(a) illustrates
this property. However, for larger $|m|$, $\Pi$ deviates from $\Lambda $
for most values of $J$, as the data for $m=-0.6$ show in 
Fig.~\ref{fig:Fig8}(b). To emphasize the advantage of $\Pi $ over $\Lambda $ as a measure of
polarization, we note that there is a regime when $\Lambda $ remains
positive even for heterophilic systems ($J<0$)! By contrast, the sign of 
$\Pi $ alone indicates which type of bias the agents have. We chose the
parameters $(m,J)=(-0.6,-0.1)$ in the run for Fig.~\ref{fig:Fig9}(a), for which 
numerically estimated $(\Lambda, \Pi) \approx (0.28, -0.10)$,
to highlight this
difference: 
while $\Lambda \approx 0.28$ implies the system is polarized, 
$\Pi \approx -0.10$ and the typical configuration clearly indicates 
{\it anti}polarization [see inset of Fig.~\ref{fig:Fig9}(a)]. Comparing the values for $\Pi$ in 
Fig.~\ref{fig:Fig8}, we note that the overall dependence on $m$ is relatively modest, which we interpret as an indication of the robustness of this measure. In
other words, being mostly free from the influence from asymmetric community sizes, 
$\Pi$ is indeed a better indicator of the effects of homophily on
polarization. In Ref.~\cite{companion}, this analysis is corroborated by mean-field results allowing us to accurately reproduce the properties of $\Lambda$ and $\Pi$ for arbitrary $m$ and $J$. 

In Fig.~\ref{fig:Fig9}(b), we illustrate the polarization and joint degree distribution under high heterogeneity and large asymmetry, $(m,J)=(-0.6,-0.6)$, in which case we have computed $(\Lambda, \Pi)\approx( -0.55, -0.87)$, corresponding to a high level of antipolarization.  As discussed above, in the case of large asymmetry and high level of heterophily, the minority agents are ``overwhelmed'' by the majority, and their degree distribution is Gaussian-like with a mean much larger than $\kappa$; see Fig.~\ref{fig:Fig7}(b). This phenomenon is also clearly noticeable in Fig.~\ref{fig:Fig9}(b), where $P_+$ differs greatly from  $P_-$ and from the joint distributions of Fig.~\ref{fig:Fig9}(a), and nearly all minority agents have CLs and a degree exceeding $\kappa$.
The comparison of Figs.~\ref{fig:Fig7}(b) and~\ref{fig:Fig9}(b) gives an insight into finite-size effects: while all the minority agents are cutters and there are no ILs within the minority community when $(N,\kappa)=(1000, 60.5)$ [see also Fig.~\ref{fig:Fig6}(b)], a small number of minority agents are cutters and have a few ILs when $(N,\kappa)=(100, 6.5)$.

\section{Discussion and Conclusion}
We have considered the dynamics of an out-of-equilibrium two-party network evolution model where agents hold fixed opinions and form dynamical links. These try to satisfy a preferred degree by endlessly creating and deleting edges. We have introduced homophily, a form of social interaction, to the simple preferred degree network dynamics. Unlike most network models with homophily~\cite{karimi2018homophily,gargiulo2017role}, here the update rules are evolutionary, and homophily (or heterophily)
influences the rate at which edges are created and removed.

Here, 
we have studied in detail systems where the parties are of the
same size using both simulation techniques and mean-field theories.
The excellent agreement between the analytical predictions and 
simulation results 
shows that we understand how the varying level of
homophily shapes the degree distribution,
the number of links across communities, and the level of polarization.  
These can help understand the phenomenon of filter bubbles~\cite{pariser2011filter} and echo chambers~\cite{barbera2015tweeting,del2016spreading,bakshy2015exposure,Iyengar2012,wang2020public}, especially when the level of polarization is high, which corresponds to the network where both parties have equally high levels of the fraction of internal links (influence assortment)~\cite{stewart2019information}, resulting in self-constructed echo chambers there.

Our model, under extreme homophily, exhibits complete fission of the population into two disconnected and polarized communities, previously found in models with rewiring~\cite{vazquez2008generic,vazquez2008analytical,durrett2012graph}.
We have also introduced an original measure of polarization that does not
share the counterintuitive properties associated with the
average edge homogeneity, commonly used in the literature, and depends weakly on the community sizes.

Simulation results, corroborated by the detailed analysis of Ref.~\cite{companion}, show that our model exhibits  a rich set of behavior when communities are of different sizes, especially 
when  moderate to high levels of heterophily are present. In particular, 
a striking feature of this model is the existence of an ``overwhelming transition'':  under sufficient heterophily, agents in the smaller group are ``overwhelmed'' by links created by members of the majority group and only try to delete edges, and their degree distribution is  Gaussian-like with an average much greater than the preferred degree. This transition therefore differs from fragmentation or fission~\cite{holme2006nonequilibrium,vazquez2008generic,durrett2012graph} and transition to paradise~\cite{gorski2020} found in other network models with homophily.
Our dynamic network model shaped by homophily therefore  appears to be generally homogeneous
with total degree distribution centered about the preferred degree, at the remarkable exception of the agents'  minority group that have a broad distribution and large degrees in the overwhelming phase.

The overwhelming transition is here attributed to the joint effect of heterophily and the existence of preferred degree.  
It would hence be interesting to investigate whether these two ingredients are  sufficient to lead to a similar transition in other models, like those of Refs.~\cite{asikainen2020cumulative,karimi2018homophily,gargiulo2017role}, and whether the  overwhelming transition is a generic feature of network models with homophily and a form of degree preference (not necessarily a strict degree value as here). It is also intriguing to notice that the increase of the degree of the minority group under sufficient heterophily,
 a salient feature of our model that is related to the  overwhelming transition, has also been 
found 
in a two-community network growing according to the preferential attachment dynamics where it originates from a different mechanism~\cite{karimi2018homophily}.
Just as rewiring schemes~\cite{lindquist2009network} can be naturally extended to co-evolutionary models, where network varies in response to changes of node states
and these change in response to updates of the network links~\cite{nardini2008s,iniguez2009opinion,gross2008adaptive,holme2006nonequilibrium,vazquez2008generic,vazquez2008analytical,durrett2012graph,henry2011emergence}, our model can be generalized to include coupled node and link co-evolutionary dynamics.  We expect that the phenomenology of such 
a co-evolutionary dynamics of a preferred degree network with homophily
will lead to an even richer and more complex phenomenology, to understand which this study is certainly a necessary building block.

\acknowledgments
We thank Andrew Mellor for substantive input, as well as Kevin
Bassler and Henk Hilhorst for helpful discussions. The support of a
joint PhD studentship of the Chinese Scholarship Council and University of
Leeds to X.L. is gratefully acknowledged (Grant No.~201803170212). We are also grateful to the London Mathematical Society (Grant No. 41712) and Leeds School of Mathematics for their financial support, and R.K.P.Z. is thankful to the Leeds School of Mathematics for their hospitality at an early stage of this collaboration. This work was partly undertaken on ARC4, part of the High Performance Computing facilities at the University of Leeds, UK.


\bibliographystyle{apsrev4-2}
\bibliography{refs}

\begin{thebibliography}{78}%
\makeatletter
\providecommand \@ifxundefined [1]{%
 \@ifx{#1\undefined}
}%
\providecommand \@ifnum [1]{%
 \ifnum #1\expandafter \@firstoftwo
 \else \expandafter \@secondoftwo
 \fi
}%
\providecommand \@ifx [1]{%
 \ifx #1\expandafter \@firstoftwo
 \else \expandafter \@secondoftwo
 \fi
}%
\providecommand \natexlab [1]{#1}%
\providecommand \enquote  [1]{``#1''}%
\providecommand \bibnamefont  [1]{#1}%
\providecommand \bibfnamefont [1]{#1}%
\providecommand \citenamefont [1]{#1}%
\providecommand \href@noop [0]{\@secondoftwo}%
\providecommand \href [0]{\begingroup \@sanitize@url \@href}%
\providecommand \@href[1]{\@@startlink{#1}\@@href}%
\providecommand \@@href[1]{\endgroup#1\@@endlink}%
\providecommand \@sanitize@url [0]{\catcode `\\12\catcode `\$12\catcode
  `\&12\catcode `\#12\catcode `\^12\catcode `\_12\catcode `\%12\relax}%
\providecommand \@@startlink[1]{}%
\providecommand \@@endlink[0]{}%
\providecommand \url  [0]{\begingroup\@sanitize@url \@url }%
\providecommand \@url [1]{\endgroup\@href {#1}{\urlprefix }}%
\providecommand \urlprefix  [0]{URL }%
\providecommand \Eprint [0]{\href }%
\providecommand \doibase [0]{https://doi.org/}%
\providecommand \selectlanguage [0]{\@gobble}%
\providecommand \bibinfo  [0]{\@secondoftwo}%
\providecommand \bibfield  [0]{\@secondoftwo}%
\providecommand \translation [1]{[#1]}%
\providecommand \BibitemOpen [0]{}%
\providecommand \bibitemStop [0]{}%
\providecommand \bibitemNoStop [0]{.\EOS\space}%
\providecommand \EOS [0]{\spacefactor3000\relax}%
\providecommand \BibitemShut  [1]{\csname bibitem#1\endcsname}%
\let\auto@bib@innerbib\@empty
\bibitem [{\citenamefont {Schelling}(1978)}]{Schelling-book}%
  \BibitemOpen
  \bibfield  {author} {\bibinfo {author} {\bibfnamefont {T.~C.}\ \bibnamefont
  {Schelling}},\ }\href@noop {} {\emph {\bibinfo {title} {Micromotives and
  Macrobehavior}}}\ (\bibinfo  {publisher} {WW Norton and Company, New York},\
  \bibinfo {year} {1978})\BibitemShut {NoStop}%
\bibitem [{\citenamefont {Castellano}\ \emph
  {et~al.}(2009{\natexlab{a}})\citenamefont {Castellano}, \citenamefont
  {Fortunato},\ and\ \citenamefont {Loreto}}]{Castellano-rev}%
  \BibitemOpen
  \bibfield  {author} {\bibinfo {author} {\bibfnamefont {C.}~\bibnamefont
  {Castellano}}, \bibinfo {author} {\bibfnamefont {S.}~\bibnamefont
  {Fortunato}},\ and\ \bibinfo {author} {\bibfnamefont {V.}~\bibnamefont
  {Loreto}},\ }\href@noop {} {\bibfield  {journal} {\bibinfo  {journal}
  {Rev.~Mod.~Phys.}\ }\textbf {\bibinfo {volume} {81}},\ \bibinfo {pages} {591}
  (\bibinfo {year} {2009}{\natexlab{a}})}\BibitemShut {NoStop}%
\bibitem [{\citenamefont {Galam}(2012)}]{Galam-book}%
  \BibitemOpen
  \bibfield  {author} {\bibinfo {author} {\bibfnamefont {S.}~\bibnamefont
  {Galam}},\ }\href@noop {} {\emph {\bibinfo {title} {Sociophysics: A
  Physicist's Modeling of Psycho-political Phenomena}}}\ (\bibinfo  {publisher}
  {Springer Science and Business Media, New York},\ \bibinfo {year}
  {2012})\BibitemShut {NoStop}%
\bibitem [{\citenamefont {Sen}\ and\ \citenamefont
  {Chakrabarti}(2013)}]{Sen-book}%
  \BibitemOpen
  \bibfield  {author} {\bibinfo {author} {\bibfnamefont {P.}~\bibnamefont
  {Sen}}\ and\ \bibinfo {author} {\bibfnamefont {B.~K.}\ \bibnamefont
  {Chakrabarti}},\ }\href@noop {} {\emph {\bibinfo {title} {Sociophysics: An
  Introduction}}}\ (\bibinfo  {publisher} {Oxford University Press, Oxford},\
  \bibinfo {year} {2013})\BibitemShut {NoStop}%
\bibitem [{\citenamefont {Jedrzejewski}\ and\ \citenamefont
  {Sznajd-Weron}(2019)}]{SW-2019}%
  \BibitemOpen
  \bibfield  {author} {\bibinfo {author} {\bibfnamefont {A.}~\bibnamefont
  {Jedrzejewski}}\ and\ \bibinfo {author} {\bibfnamefont {K.}~\bibnamefont
  {Sznajd-Weron}},\ }\href@noop {} {\bibfield  {journal} {\bibinfo  {journal}
  {C. R. Physique}\ }\textbf {\bibinfo {volume} {20}},\ \bibinfo {pages} {244}
  (\bibinfo {year} {2019})}\BibitemShut {NoStop}%
\bibitem [{\citenamefont {Albert}\ and\ \citenamefont
  {Barab\'asi}(2002)}]{Albert-rev}%
  \BibitemOpen
  \bibfield  {author} {\bibinfo {author} {\bibfnamefont {R.}~\bibnamefont
  {Albert}}\ and\ \bibinfo {author} {\bibfnamefont {L.}~\bibnamefont
  {Barab\'asi}},\ }\href@noop {} {\bibfield  {journal} {\bibinfo  {journal}
  {Rev. Mod. Phys.}\ }\textbf {\bibinfo {volume} {74}},\ \bibinfo {pages} {47}
  (\bibinfo {year} {2002})}\BibitemShut {NoStop}%
\bibitem [{\citenamefont {Dorogovtsev}\ and\ \citenamefont
  {Mendes}(2003)}]{Dorogovtsev-book}%
  \BibitemOpen
  \bibfield  {author} {\bibinfo {author} {\bibfnamefont {S.~N.}\ \bibnamefont
  {Dorogovtsev}}\ and\ \bibinfo {author} {\bibfnamefont {J.~F.~F.}\
  \bibnamefont {Mendes}},\ }\href@noop {} {\emph {\bibinfo {title} {Evolution
  of Networks: From Biological Nets to the Internet and WWW}}}\ (\bibinfo
  {publisher} {Oxford University Press, Oxford},\ \bibinfo {year}
  {2003})\BibitemShut {NoStop}%
\bibitem [{\citenamefont {Newman}(2010)}]{newman2018networks}%
  \BibitemOpen
  \bibfield  {author} {\bibinfo {author} {\bibfnamefont {M.}~\bibnamefont
  {Newman}},\ }\href@noop {} {\emph {\bibinfo {title} {Networks}}}\ (\bibinfo
  {publisher} {Oxford University Press, Oxford},\ \bibinfo {year}
  {2010})\BibitemShut {NoStop}%
\bibitem [{\citenamefont {Szabo}\ and\ \citenamefont
  {F\'ath}(2007)}]{Szabo-rev}%
  \BibitemOpen
  \bibfield  {author} {\bibinfo {author} {\bibfnamefont {G.}~\bibnamefont
  {Szabo}}\ and\ \bibinfo {author} {\bibfnamefont {G.}~\bibnamefont {F\'ath}},\
  }\href@noop {} {\bibfield  {journal} {\bibinfo  {journal} {Phys. Rep.}\
  }\textbf {\bibinfo {volume} {446}},\ \bibinfo {pages} {97} (\bibinfo {year}
  {2007})}\BibitemShut {NoStop}%
\bibitem [{\citenamefont {Antal}\ \emph {et~al.}(2006)\citenamefont {Antal},
  \citenamefont {Redner},\ and\ \citenamefont {Sood}}]{Antal-2006}%
  \BibitemOpen
  \bibfield  {author} {\bibinfo {author} {\bibfnamefont {T.}~\bibnamefont
  {Antal}}, \bibinfo {author} {\bibfnamefont {S.}~\bibnamefont {Redner}},\ and\
  \bibinfo {author} {\bibfnamefont {V.}~\bibnamefont {Sood}},\ }\href@noop {}
  {\bibfield  {journal} {\bibinfo  {journal} {Phys. Rev. Lett.}\ }\textbf
  {\bibinfo {volume} {96}},\ \bibinfo {pages} {188104} (\bibinfo {year}
  {2006})}\BibitemShut {NoStop}%
\bibitem [{\citenamefont {Sood}\ \emph {et~al.}(2008)\citenamefont {Sood},
  \citenamefont {Antal},\ and\ \citenamefont {Redner}}]{sood2008voter}%
  \BibitemOpen
  \bibfield  {author} {\bibinfo {author} {\bibfnamefont {V.}~\bibnamefont
  {Sood}}, \bibinfo {author} {\bibfnamefont {T.}~\bibnamefont {Antal}},\ and\
  \bibinfo {author} {\bibfnamefont {S.}~\bibnamefont {Redner}},\ }\href@noop {}
  {\bibfield  {journal} {\bibinfo  {journal} {Phys.~Rev.~E}\ }\textbf {\bibinfo
  {volume} {77}},\ \bibinfo {pages} {041121} (\bibinfo {year}
  {2008})}\BibitemShut {NoStop}%
\bibitem [{\citenamefont {Baxter}\ \emph {et~al.}(2008)\citenamefont {Baxter},
  \citenamefont {Blythe},\ and\ \citenamefont {McKane}}]{Baxter-2008}%
  \BibitemOpen
  \bibfield  {author} {\bibinfo {author} {\bibfnamefont {G.~J.}\ \bibnamefont
  {Baxter}}, \bibinfo {author} {\bibfnamefont {R.~A.}\ \bibnamefont {Blythe}},\
  and\ \bibinfo {author} {\bibfnamefont {A.~J.}\ \bibnamefont {McKane}},\
  }\href@noop {} {\bibfield  {journal} {\bibinfo  {journal} {Phys. Rev. Lett.}\
  }\textbf {\bibinfo {volume} {101}},\ \bibinfo {pages} {258701} (\bibinfo
  {year} {2008})}\BibitemShut {NoStop}%
\bibitem [{\citenamefont {Blythe}(2010)}]{Blythe-2010}%
  \BibitemOpen
  \bibfield  {author} {\bibinfo {author} {\bibfnamefont {R.~A.}\ \bibnamefont
  {Blythe}},\ }\href@noop {} {\bibfield  {journal} {\bibinfo  {journal} {J.
  Phys. A: Math. Theor.}\ }\textbf {\bibinfo {volume} {43}},\ \bibinfo {pages}
  {385003} (\bibinfo {year} {2010})}\BibitemShut {NoStop}%
\bibitem [{\citenamefont {Castellano}\ \emph {et~al.}(2000)\citenamefont
  {Castellano}, \citenamefont {Marsili},\ and\ \citenamefont
  {Vespignani}}]{Castellano-2010}%
  \BibitemOpen
  \bibfield  {author} {\bibinfo {author} {\bibfnamefont {C.}~\bibnamefont
  {Castellano}}, \bibinfo {author} {\bibfnamefont {M.}~\bibnamefont
  {Marsili}},\ and\ \bibinfo {author} {\bibfnamefont {A.}~\bibnamefont
  {Vespignani}},\ }\href@noop {} {\bibfield  {journal} {\bibinfo  {journal}
  {Phys.~Rev.~Lett.}\ }\textbf {\bibinfo {volume} {85}},\ \bibinfo {pages}
  {3536} (\bibinfo {year} {2000})}\BibitemShut {NoStop}%
\bibitem [{\citenamefont {Moretti}\ \emph {et~al.}(2013)\citenamefont
  {Moretti}, \citenamefont {Liu}, \citenamefont {Castellano},\ and\
  \citenamefont {Pastor-Satorras}}]{moretti2013mean}%
  \BibitemOpen
  \bibfield  {author} {\bibinfo {author} {\bibfnamefont {P.}~\bibnamefont
  {Moretti}}, \bibinfo {author} {\bibfnamefont {S.}~\bibnamefont {Liu}},
  \bibinfo {author} {\bibfnamefont {C.}~\bibnamefont {Castellano}},\ and\
  \bibinfo {author} {\bibfnamefont {R.}~\bibnamefont {Pastor-Satorras}},\
  }\href@noop {} {\bibfield  {journal} {\bibinfo  {journal} {J. Stat. Phys.}\
  }\textbf {\bibinfo {volume} {151}},\ \bibinfo {pages} {113} (\bibinfo {year}
  {2013})}\BibitemShut {NoStop}%
\bibitem [{\citenamefont {Szolnoki}\ \emph {et~al.}(2014)\citenamefont
  {Szolnoki}, \citenamefont {Perc},\ and\ \citenamefont
  {Mobilia}}]{Szolnoki-2014}%
  \BibitemOpen
  \bibfield  {author} {\bibinfo {author} {\bibfnamefont {A.}~\bibnamefont
  {Szolnoki}}, \bibinfo {author} {\bibfnamefont {M.}~\bibnamefont {Perc}},\
  and\ \bibinfo {author} {\bibfnamefont {M.}~\bibnamefont {Mobilia}},\
  }\href@noop {} {\bibfield  {journal} {\bibinfo  {journal} {Phys.~Rev.~E}\
  }\textbf {\bibinfo {volume} {89}},\ \bibinfo {pages} {042802} (\bibinfo
  {year} {2014})}\BibitemShut {NoStop}%
\bibitem [{\citenamefont {Sabsovich}\ \emph {et~al.}(2017)\citenamefont
  {Sabsovich}, \citenamefont {Mobilia},\ and\ \citenamefont
  {Assaf}}]{Sabsovich-2017}%
  \BibitemOpen
  \bibfield  {author} {\bibinfo {author} {\bibfnamefont {D.}~\bibnamefont
  {Sabsovich}}, \bibinfo {author} {\bibfnamefont {M.}~\bibnamefont {Mobilia}},\
  and\ \bibinfo {author} {\bibfnamefont {M.}~\bibnamefont {Assaf}},\
  }\href@noop {} {\bibfield  {journal} {\bibinfo  {journal} {J. Stat. Mech.:
  Theory Exp.}\ }\textbf {\bibinfo {volume} {2017}},\ \bibinfo {pages}
  {053405}}\BibitemShut {NoStop}%
\bibitem [{\citenamefont {Newman}(2003)}]{Newman-2003}%
  \BibitemOpen
  \bibfield  {author} {\bibinfo {author} {\bibfnamefont {M.~E.~J.}\
  \bibnamefont {Newman}},\ }\href@noop {} {\bibfield  {journal} {\bibinfo
  {journal} {Phys.~Rev.~E}\ }\textbf {\bibinfo {volume} {67}},\ \bibinfo
  {pages} {026126} (\bibinfo {year} {2003})}\BibitemShut {NoStop}%
\bibitem [{\citenamefont {Bogun\'a}\ \emph {et~al.}(2004)\citenamefont
  {Bogun\'a}, \citenamefont {Pastor-Satorras}, \citenamefont {Diaz-Guilera},\
  and\ \citenamefont {Arenas}}]{Boguna-2004}%
  \BibitemOpen
  \bibfield  {author} {\bibinfo {author} {\bibfnamefont {M.}~\bibnamefont
  {Bogun\'a}}, \bibinfo {author} {\bibfnamefont {R.}~\bibnamefont
  {Pastor-Satorras}}, \bibinfo {author} {\bibfnamefont {A.}~\bibnamefont
  {Diaz-Guilera}},\ and\ \bibinfo {author} {\bibfnamefont {A.}~\bibnamefont
  {Arenas}},\ }\href@noop {} {\bibfield  {journal} {\bibinfo  {journal}
  {Phys.~Rev.~E}\ }\textbf {\bibinfo {volume} {70}},\ \bibinfo {pages} {056122}
  (\bibinfo {year} {2004})}\BibitemShut {NoStop}%
\bibitem [{\citenamefont {Murase}\ \emph {et~al.}(2019)\citenamefont {Murase},
  \citenamefont {Jo}, \citenamefont {T\"or\"ok}, \citenamefont {Kert\'esz},\
  and\ \citenamefont {Kaski}}]{Murase-2019}%
  \BibitemOpen
  \bibfield  {author} {\bibinfo {author} {\bibfnamefont {Y.}~\bibnamefont
  {Murase}}, \bibinfo {author} {\bibfnamefont {H.-H.}\ \bibnamefont {Jo}},
  \bibinfo {author} {\bibfnamefont {J.}~\bibnamefont {T\"or\"ok}}, \bibinfo
  {author} {\bibfnamefont {J.}~\bibnamefont {Kert\'esz}},\ and\ \bibinfo
  {author} {\bibfnamefont {K.}~\bibnamefont {Kaski}},\ }\href@noop {}
  {\bibfield  {journal} {\bibinfo  {journal} {Sci.~Rep.}\ }\textbf {\bibinfo
  {volume} {9}},\ \bibinfo {pages} {4310} (\bibinfo {year} {2019})}\BibitemShut
  {NoStop}%
\bibitem [{\citenamefont {McPherson}\ \emph {et~al.}(2001)\citenamefont
  {McPherson}, \citenamefont {Smith-Lovin},\ and\ \citenamefont
  {Cook}}]{mcpherson2001birds}%
  \BibitemOpen
  \bibfield  {author} {\bibinfo {author} {\bibfnamefont {M.}~\bibnamefont
  {McPherson}}, \bibinfo {author} {\bibfnamefont {L.}~\bibnamefont
  {Smith-Lovin}},\ and\ \bibinfo {author} {\bibfnamefont {J.~M.}\ \bibnamefont
  {Cook}},\ }\href@noop {} {\bibfield  {journal} {\bibinfo  {journal} {Ann.
  Rev. Sociol.}\ }\textbf {\bibinfo {volume} {27}},\ \bibinfo {pages} {415}
  (\bibinfo {year} {2001})}\BibitemShut {NoStop}%
\bibitem [{\citenamefont {Centola}(2011)}]{centola2011experimental}%
  \BibitemOpen
  \bibfield  {author} {\bibinfo {author} {\bibfnamefont {D.}~\bibnamefont
  {Centola}},\ }\href@noop {} {\bibfield  {journal} {\bibinfo  {journal}
  {Science}\ }\textbf {\bibinfo {volume} {334}},\ \bibinfo {pages} {1269}
  (\bibinfo {year} {2011})}\BibitemShut {NoStop}%
\bibitem [{\citenamefont {Centola}\ and\ \citenamefont
  {Macy}(2007)}]{centola2007complex}%
  \BibitemOpen
  \bibfield  {author} {\bibinfo {author} {\bibfnamefont {D.}~\bibnamefont
  {Centola}}\ and\ \bibinfo {author} {\bibfnamefont {M.}~\bibnamefont {Macy}},\
  }\href@noop {} {\bibfield  {journal} {\bibinfo  {journal} {Am. J. Sociol.}\
  }\textbf {\bibinfo {volume} {113}},\ \bibinfo {pages} {702} (\bibinfo {year}
  {2007})}\BibitemShut {NoStop}%
\bibitem [{\citenamefont {Del~Vicario}\ \emph {et~al.}(2017)\citenamefont
  {Del~Vicario}, \citenamefont {Scala}, \citenamefont {Caldarelli},
  \citenamefont {Stanley},\ and\ \citenamefont
  {Quattrociocchi}}]{del2017modeling}%
  \BibitemOpen
  \bibfield  {author} {\bibinfo {author} {\bibfnamefont {M.}~\bibnamefont
  {Del~Vicario}}, \bibinfo {author} {\bibfnamefont {A.}~\bibnamefont {Scala}},
  \bibinfo {author} {\bibfnamefont {G.}~\bibnamefont {Caldarelli}}, \bibinfo
  {author} {\bibfnamefont {H.~E.}\ \bibnamefont {Stanley}},\ and\ \bibinfo
  {author} {\bibfnamefont {W.}~\bibnamefont {Quattrociocchi}},\ }\href@noop {}
  {\bibfield  {journal} {\bibinfo  {journal} {Sci.~Rep.}\ }\textbf {\bibinfo
  {volume} {7}},\ \bibinfo {pages} {40391} (\bibinfo {year}
  {2017})}\BibitemShut {NoStop}%
\bibitem [{\citenamefont {Centola}\ \emph {et~al.}(2007)\citenamefont
  {Centola}, \citenamefont {Gonzalez-Avella}, \citenamefont {Eguiluz},\ and\
  \citenamefont {Miguel}}]{centola2007homophily}%
  \BibitemOpen
  \bibfield  {author} {\bibinfo {author} {\bibfnamefont {D.}~\bibnamefont
  {Centola}}, \bibinfo {author} {\bibfnamefont {J.~C.}\ \bibnamefont
  {Gonzalez-Avella}}, \bibinfo {author} {\bibfnamefont {V.~M.}\ \bibnamefont
  {Eguiluz}},\ and\ \bibinfo {author} {\bibfnamefont {M.~S.}\ \bibnamefont
  {Miguel}},\ }\href@noop {} {\bibfield  {journal} {\bibinfo  {journal} {J.
  Confl. Resolut.}\ }\textbf {\bibinfo {volume} {51}},\ \bibinfo {pages} {905}
  (\bibinfo {year} {2007})}\BibitemShut {NoStop}%
\bibitem [{\citenamefont {Volkovich}\ \emph {et~al.}(2014)\citenamefont
  {Volkovich}, \citenamefont {Laniado}, \citenamefont {Kappler},\ and\
  \citenamefont {Kaltenbrunner}}]{volkovich2014}%
  \BibitemOpen
  \bibfield  {author} {\bibinfo {author} {\bibfnamefont {Y.}~\bibnamefont
  {Volkovich}}, \bibinfo {author} {\bibfnamefont {D.}~\bibnamefont {Laniado}},
  \bibinfo {author} {\bibfnamefont {K.~E.}\ \bibnamefont {Kappler}},\ and\
  \bibinfo {author} {\bibfnamefont {A.}~\bibnamefont {Kaltenbrunner}},\ }in\
  \href@noop {} {\emph {\bibinfo {booktitle} {International Conference on
  Social Informatics}}}\ (\bibinfo {organization} {Springer, Berlin},\ \bibinfo
  {year} {2014})\ pp.\ \bibinfo {pages} {139--150}\BibitemShut {NoStop}%
\bibitem [{\citenamefont {Zeltzer}(2020)}]{zeltzer2020gender}%
  \BibitemOpen
  \bibfield  {author} {\bibinfo {author} {\bibfnamefont {D.}~\bibnamefont
  {Zeltzer}},\ }\href@noop {} {\bibfield  {journal} {\bibinfo  {journal} {Am.
  Econ. J.: Appl. Econ.}\ }\textbf {\bibinfo {volume} {12}},\ \bibinfo {pages}
  {169} (\bibinfo {year} {2020})}\BibitemShut {NoStop}%
\bibitem [{\citenamefont {McPherson}\ and\ \citenamefont
  {Smith-Lovin}(1987)}]{mcpherson1987homophily}%
  \BibitemOpen
  \bibfield  {author} {\bibinfo {author} {\bibfnamefont {J.~M.}\ \bibnamefont
  {McPherson}}\ and\ \bibinfo {author} {\bibfnamefont {L.}~\bibnamefont
  {Smith-Lovin}},\ }\href@noop {} {\bibfield  {journal} {\bibinfo  {journal}
  {Am. Sociol. Rev.}\ }\textbf {\bibinfo {volume} {52}},\ \bibinfo {pages}
  {370} (\bibinfo {year} {1987})}\BibitemShut {NoStop}%
\bibitem [{\citenamefont {Yava{\c{s}}}\ and\ \citenamefont
  {Y{\"u}cel}(2014)}]{yavacs2014impact}%
  \BibitemOpen
  \bibfield  {author} {\bibinfo {author} {\bibfnamefont {M.}~\bibnamefont
  {Yava{\c{s}}}}\ and\ \bibinfo {author} {\bibfnamefont {G.}~\bibnamefont
  {Y{\"u}cel}},\ }\href@noop {} {\bibfield  {journal} {\bibinfo  {journal}
  {Soc. Sci. Comput. Rev.}\ }\textbf {\bibinfo {volume} {32}},\ \bibinfo
  {pages} {354} (\bibinfo {year} {2014})}\BibitemShut {NoStop}%
\bibitem [{\citenamefont {Shalizi}\ and\ \citenamefont
  {Thomas}(2011)}]{shalizi2011homophily}%
  \BibitemOpen
  \bibfield  {author} {\bibinfo {author} {\bibfnamefont {C.~R.}\ \bibnamefont
  {Shalizi}}\ and\ \bibinfo {author} {\bibfnamefont {A.~C.}\ \bibnamefont
  {Thomas}},\ }\href@noop {} {\bibfield  {journal} {\bibinfo  {journal}
  {Sociol. Method Res.}\ }\textbf {\bibinfo {volume} {40}},\ \bibinfo {pages}
  {211} (\bibinfo {year} {2011})}\BibitemShut {NoStop}%
\bibitem [{\citenamefont {Pariser}(2011)}]{pariser2011filter}%
  \BibitemOpen
  \bibfield  {author} {\bibinfo {author} {\bibfnamefont {E.}~\bibnamefont
  {Pariser}},\ }\href@noop {} {\emph {\bibinfo {title} {The filter bubble: What
  the Internet is Hiding from You}}}\ (\bibinfo  {publisher} {Penguin London},\
  \bibinfo {year} {2011})\BibitemShut {NoStop}%
\bibitem [{\citenamefont {Iyengar}\ \emph {et~al.}(2012)\citenamefont
  {Iyengar}, \citenamefont {Sood},\ and\ \citenamefont {Lelkes}}]{Iyengar2012}%
  \BibitemOpen
  \bibfield  {author} {\bibinfo {author} {\bibfnamefont {S.}~\bibnamefont
  {Iyengar}}, \bibinfo {author} {\bibfnamefont {G.}~\bibnamefont {Sood}},\ and\
  \bibinfo {author} {\bibfnamefont {Y.}~\bibnamefont {Lelkes}},\ }\href@noop {}
  {\bibfield  {journal} {\bibinfo  {journal} {Public Opin. Q.}\ }\textbf
  {\bibinfo {volume} {76}},\ \bibinfo {pages} {405} (\bibinfo {year}
  {2012})}\BibitemShut {NoStop}%
\bibitem [{\citenamefont {Barber{\'a}}\ \emph {et~al.}(2015)\citenamefont
  {Barber{\'a}}, \citenamefont {Jost}, \citenamefont {Nagler}, \citenamefont
  {Tucker},\ and\ \citenamefont {Bonneau}}]{barbera2015tweeting}%
  \BibitemOpen
  \bibfield  {author} {\bibinfo {author} {\bibfnamefont {P.}~\bibnamefont
  {Barber{\'a}}}, \bibinfo {author} {\bibfnamefont {J.~T.}\ \bibnamefont
  {Jost}}, \bibinfo {author} {\bibfnamefont {J.}~\bibnamefont {Nagler}},
  \bibinfo {author} {\bibfnamefont {J.~A.}\ \bibnamefont {Tucker}},\ and\
  \bibinfo {author} {\bibfnamefont {R.}~\bibnamefont {Bonneau}},\ }\href@noop
  {} {\bibfield  {journal} {\bibinfo  {journal} {Psychol. Sci.}\ }\textbf
  {\bibinfo {volume} {26}},\ \bibinfo {pages} {1531} (\bibinfo {year}
  {2015})}\BibitemShut {NoStop}%
\bibitem [{\citenamefont {Barber\'a}(2015)}]{barbera2015}%
  \BibitemOpen
  \bibfield  {author} {\bibinfo {author} {\bibfnamefont {P.}~\bibnamefont
  {Barber\'a}},\ }\href@noop {} {\bibfield  {journal} {\bibinfo  {journal}
  {Political Analysis}\ }\textbf {\bibinfo {volume} {23}},\ \bibinfo {pages}
  {76} (\bibinfo {year} {2015})}\BibitemShut {NoStop}%
\bibitem [{\citenamefont {Bakshy}\ \emph {et~al.}(2015)\citenamefont {Bakshy},
  \citenamefont {Messing},\ and\ \citenamefont {Adamic}}]{bakshy2015exposure}%
  \BibitemOpen
  \bibfield  {author} {\bibinfo {author} {\bibfnamefont {E.}~\bibnamefont
  {Bakshy}}, \bibinfo {author} {\bibfnamefont {S.}~\bibnamefont {Messing}},\
  and\ \bibinfo {author} {\bibfnamefont {L.~A.}\ \bibnamefont {Adamic}},\
  }\href@noop {} {\bibfield  {journal} {\bibinfo  {journal} {Science}\ }\textbf
  {\bibinfo {volume} {348}},\ \bibinfo {pages} {1130} (\bibinfo {year}
  {2015})}\BibitemShut {NoStop}%
\bibitem [{\citenamefont {Del~Vicario}\ \emph {et~al.}(2016)\citenamefont
  {Del~Vicario}, \citenamefont {Bessi}, \citenamefont {Zollo}, \citenamefont
  {Petroni}, \citenamefont {Scala}, \citenamefont {Caldarelli}, \citenamefont
  {Stanley},\ and\ \citenamefont {Quattrociocchi}}]{del2016spreading}%
  \BibitemOpen
  \bibfield  {author} {\bibinfo {author} {\bibfnamefont {M.}~\bibnamefont
  {Del~Vicario}}, \bibinfo {author} {\bibfnamefont {A.}~\bibnamefont {Bessi}},
  \bibinfo {author} {\bibfnamefont {F.}~\bibnamefont {Zollo}}, \bibinfo
  {author} {\bibfnamefont {F.}~\bibnamefont {Petroni}}, \bibinfo {author}
  {\bibfnamefont {A.}~\bibnamefont {Scala}}, \bibinfo {author} {\bibfnamefont
  {G.}~\bibnamefont {Caldarelli}}, \bibinfo {author} {\bibfnamefont {H.~E.}\
  \bibnamefont {Stanley}},\ and\ \bibinfo {author} {\bibfnamefont
  {W.}~\bibnamefont {Quattrociocchi}},\ }\href@noop {} {\bibfield  {journal}
  {\bibinfo  {journal} {Proc. Natl. Acad. Sci. U.S.A.}\ }\textbf {\bibinfo
  {volume} {113}},\ \bibinfo {pages} {554} (\bibinfo {year}
  {2016})}\BibitemShut {NoStop}%
\bibitem [{\citenamefont {Wang}\ \emph {et~al.}(2020)\citenamefont {Wang},
  \citenamefont {Sirianni}, \citenamefont {Tang}, \citenamefont {Zheng},\ and\
  \citenamefont {Fu}}]{wang2020public}%
  \BibitemOpen
  \bibfield  {author} {\bibinfo {author} {\bibfnamefont {X.}~\bibnamefont
  {Wang}}, \bibinfo {author} {\bibfnamefont {A.~D.}\ \bibnamefont {Sirianni}},
  \bibinfo {author} {\bibfnamefont {S.}~\bibnamefont {Tang}}, \bibinfo {author}
  {\bibfnamefont {Z.}~\bibnamefont {Zheng}},\ and\ \bibinfo {author}
  {\bibfnamefont {F.}~\bibnamefont {Fu}},\ }\href@noop {} {\bibfield  {journal}
  {\bibinfo  {journal} {Phys. Rev. X}\ }\textbf {\bibinfo {volume} {10}},\
  \bibinfo {pages} {041042} (\bibinfo {year} {2020})}\BibitemShut {NoStop}%
\bibitem [{\citenamefont {Xie}\ \emph {et~al.}(2016)\citenamefont {Xie},
  \citenamefont {Li}, \citenamefont {Jiang}, \citenamefont {Tan}, \citenamefont
  {Podobnik}, \citenamefont {Zhou},\ and\ \citenamefont {Stanley}}]{xie2016}%
  \BibitemOpen
  \bibfield  {author} {\bibinfo {author} {\bibfnamefont {W.}~\bibnamefont
  {Xie}}, \bibinfo {author} {\bibfnamefont {M.-X.}\ \bibnamefont {Li}},
  \bibinfo {author} {\bibfnamefont {Z.-Q.}\ \bibnamefont {Jiang}}, \bibinfo
  {author} {\bibfnamefont {Q.-Z.}\ \bibnamefont {Tan}}, \bibinfo {author}
  {\bibfnamefont {B.}~\bibnamefont {Podobnik}}, \bibinfo {author}
  {\bibfnamefont {W.-X.}\ \bibnamefont {Zhou}},\ and\ \bibinfo {author}
  {\bibfnamefont {H.~E.}\ \bibnamefont {Stanley}},\ }\href@noop {} {\bibfield
  {journal} {\bibinfo  {journal} {Sci.~Rep.}\ }\textbf {\bibinfo {volume}
  {6}},\ \bibinfo {pages} {18727} (\bibinfo {year} {2016})}\BibitemShut
  {NoStop}%
\bibitem [{\citenamefont {Ramazi}\ \emph {et~al.}(2018)\citenamefont {Ramazi},
  \citenamefont {Riehl},\ and\ \citenamefont {M.}}]{ramazi2018}%
  \BibitemOpen
  \bibfield  {author} {\bibinfo {author} {\bibfnamefont {P.}~\bibnamefont
  {Ramazi}}, \bibinfo {author} {\bibfnamefont {J.}~\bibnamefont {Riehl}},\ and\
  \bibinfo {author} {\bibfnamefont {C.}~\bibnamefont {M.}},\ }\href@noop {}
  {\bibfield  {journal} {\bibinfo  {journal} {R. Soc. Open Sci.}\ }\textbf
  {\bibinfo {volume} {5}},\ \bibinfo {pages} {180027} (\bibinfo {year}
  {2018})}\BibitemShut {NoStop}%
\bibitem [{\citenamefont {Barranco}\ \emph {et~al.}(2019)\citenamefont
  {Barranco}, \citenamefont {Lozares},\ and\ \citenamefont
  {Muntanyola‐Saura}}]{barranco2019}%
  \BibitemOpen
  \bibfield  {author} {\bibinfo {author} {\bibfnamefont {O.}~\bibnamefont
  {Barranco}}, \bibinfo {author} {\bibfnamefont {C.}~\bibnamefont {Lozares}},\
  and\ \bibinfo {author} {\bibfnamefont {D.}~\bibnamefont
  {Muntanyola‐Saura}},\ }\href@noop {} {\bibfield  {journal} {\bibinfo
  {journal} {Qual. Quant.}\ }\textbf {\bibinfo {volume} {53}},\ \bibinfo
  {pages} {599–619} (\bibinfo {year} {2019})}\BibitemShut {NoStop}%
\bibitem [{\citenamefont {Yokomatsu}\ and\ \citenamefont
  {Kotani}(2021)}]{yokomatsu2021}%
  \BibitemOpen
  \bibfield  {author} {\bibinfo {author} {\bibfnamefont {M.}~\bibnamefont
  {Yokomatsu}}\ and\ \bibinfo {author} {\bibfnamefont {H.}~\bibnamefont
  {Kotani}},\ }\href@noop {} {\bibfield  {journal} {\bibinfo  {journal} {J.
  Math. Sociol.}\ }\textbf {\bibinfo {volume} {45}},\ \bibinfo {pages} {111}
  (\bibinfo {year} {2021})}\BibitemShut {NoStop}%
\bibitem [{\citenamefont {Gargiulo}\ and\ \citenamefont
  {Gandica}(2017)}]{gargiulo2017role}%
  \BibitemOpen
  \bibfield  {author} {\bibinfo {author} {\bibfnamefont {F.}~\bibnamefont
  {Gargiulo}}\ and\ \bibinfo {author} {\bibfnamefont {Y.}~\bibnamefont
  {Gandica}},\ }\href@noop {} {\bibfield  {journal} {\bibinfo  {journal} {J.
  Artific. Soc. Soc. Simul.}\ }\textbf {\bibinfo {volume} {20}},\ \bibinfo
  {pages} {8} (\bibinfo {year} {2017})}\BibitemShut {NoStop}%
\bibitem [{\citenamefont {Wong}\ \emph {et~al.}(2006)\citenamefont {Wong},
  \citenamefont {Pattison},\ and\ \citenamefont {Robins}}]{wong2006spatial}%
  \BibitemOpen
  \bibfield  {author} {\bibinfo {author} {\bibfnamefont {L.~H.}\ \bibnamefont
  {Wong}}, \bibinfo {author} {\bibfnamefont {P.}~\bibnamefont {Pattison}},\
  and\ \bibinfo {author} {\bibfnamefont {G.}~\bibnamefont {Robins}},\
  }\href@noop {} {\bibfield  {journal} {\bibinfo  {journal} {Physica A}\
  }\textbf {\bibinfo {volume} {360}},\ \bibinfo {pages} {99} (\bibinfo {year}
  {2006})}\BibitemShut {NoStop}%
\bibitem [{\citenamefont {Karimi}\ \emph {et~al.}(2018)\citenamefont {Karimi},
  \citenamefont {G{\'e}nois}, \citenamefont {Wagner}, \citenamefont {Singer},\
  and\ \citenamefont {Strohmaier}}]{karimi2018homophily}%
  \BibitemOpen
  \bibfield  {author} {\bibinfo {author} {\bibfnamefont {F.}~\bibnamefont
  {Karimi}}, \bibinfo {author} {\bibfnamefont {M.}~\bibnamefont {G{\'e}nois}},
  \bibinfo {author} {\bibfnamefont {C.}~\bibnamefont {Wagner}}, \bibinfo
  {author} {\bibfnamefont {P.}~\bibnamefont {Singer}},\ and\ \bibinfo {author}
  {\bibfnamefont {M.}~\bibnamefont {Strohmaier}},\ }\href@noop {} {\bibfield
  {journal} {\bibinfo  {journal} {Sci. Rep.}\ }\textbf {\bibinfo {volume}
  {8}},\ \bibinfo {pages} {11077} (\bibinfo {year} {2018})}\BibitemShut
  {NoStop}%
\bibitem [{\citenamefont {Kimura}\ and\ \citenamefont
  {Hayakawa}(2008)}]{kimura2008coevolutionary}%
  \BibitemOpen
  \bibfield  {author} {\bibinfo {author} {\bibfnamefont {D.}~\bibnamefont
  {Kimura}}\ and\ \bibinfo {author} {\bibfnamefont {Y.}~\bibnamefont
  {Hayakawa}},\ }\href@noop {} {\bibfield  {journal} {\bibinfo  {journal}
  {Phys. Rev. E}\ }\textbf {\bibinfo {volume} {78}},\ \bibinfo {pages} {016103}
  (\bibinfo {year} {2008})}\BibitemShut {NoStop}%
\bibitem [{\citenamefont {Papadopoulos}\ \emph {et~al.}(2012)\citenamefont
  {Papadopoulos}, \citenamefont {Kitsak}, \citenamefont {Serrano},
  \citenamefont {Bogun{\'a}},\ and\ \citenamefont
  {Krioukov}}]{papadopoulos2012popularity}%
  \BibitemOpen
  \bibfield  {author} {\bibinfo {author} {\bibfnamefont {F.}~\bibnamefont
  {Papadopoulos}}, \bibinfo {author} {\bibfnamefont {M.}~\bibnamefont
  {Kitsak}}, \bibinfo {author} {\bibfnamefont {M.~{\'A}.}\ \bibnamefont
  {Serrano}}, \bibinfo {author} {\bibfnamefont {M.}~\bibnamefont
  {Bogun{\'a}}},\ and\ \bibinfo {author} {\bibfnamefont {D.}~\bibnamefont
  {Krioukov}},\ }\href@noop {} {\bibfield  {journal} {\bibinfo  {journal}
  {Nature}\ }\textbf {\bibinfo {volume} {489}},\ \bibinfo {pages} {537}
  (\bibinfo {year} {2012})}\BibitemShut {NoStop}%
\bibitem [{\citenamefont {Asikainen}\ \emph {et~al.}(2020)\citenamefont
  {Asikainen}, \citenamefont {I{\~n}iguez}, \citenamefont
  {Ure{\~n}a-Carri{\'o}n}, \citenamefont {Kaski},\ and\ \citenamefont
  {Kivel{\"a}}}]{asikainen2020cumulative}%
  \BibitemOpen
  \bibfield  {author} {\bibinfo {author} {\bibfnamefont {A.}~\bibnamefont
  {Asikainen}}, \bibinfo {author} {\bibfnamefont {G.}~\bibnamefont
  {I{\~n}iguez}}, \bibinfo {author} {\bibfnamefont {J.}~\bibnamefont
  {Ure{\~n}a-Carri{\'o}n}}, \bibinfo {author} {\bibfnamefont {K.}~\bibnamefont
  {Kaski}},\ and\ \bibinfo {author} {\bibfnamefont {M.}~\bibnamefont
  {Kivel{\"a}}},\ }\href@noop {} {\bibfield  {journal} {\bibinfo  {journal}
  {Sci. Adv.}\ }\textbf {\bibinfo {volume} {6}},\ \bibinfo {pages} {eaax7310}
  (\bibinfo {year} {2020})}\BibitemShut {NoStop}%
\bibitem [{\citenamefont {Krapivsky}\ and\ \citenamefont
  {Redner}(2021)}]{krapivsky2021divergence}%
  \BibitemOpen
  \bibfield  {author} {\bibinfo {author} {\bibfnamefont {P.~L.}\ \bibnamefont
  {Krapivsky}}\ and\ \bibinfo {author} {\bibfnamefont {S.}~\bibnamefont
  {Redner}},\ }\href@noop {} {\bibfield  {journal} {\bibinfo  {journal}
  {Phys.~Rev.~E}\ }\textbf {\bibinfo {volume} {103}},\ \bibinfo {pages}
  {L060301} (\bibinfo {year} {2021})}\BibitemShut {NoStop}%
\bibitem [{\citenamefont {Gorski}\ \emph {et~al.}(2020)\citenamefont {Gorski},
  \citenamefont {Bochenina}, \citenamefont {Holyst},\ and\ \citenamefont
  {D'Souza}}]{gorski2020}%
  \BibitemOpen
  \bibfield  {author} {\bibinfo {author} {\bibfnamefont {P.~J.}\ \bibnamefont
  {Gorski}}, \bibinfo {author} {\bibfnamefont {K.}~\bibnamefont {Bochenina}},
  \bibinfo {author} {\bibfnamefont {J.~A.}\ \bibnamefont {Holyst}},\ and\
  \bibinfo {author} {\bibfnamefont {R.~M.}\ \bibnamefont {D'Souza}},\
  }\href@noop {} {\bibfield  {journal} {\bibinfo  {journal} {Phys.~Rev.~Lett.}\
  }\textbf {\bibinfo {volume} {125}},\ \bibinfo {pages} {078302} (\bibinfo
  {year} {2020})}\BibitemShut {NoStop}%
\bibitem [{\citenamefont {Overgoor}\ \emph {et~al.}(2019)\citenamefont
  {Overgoor}, \citenamefont {Benson},\ and\ \citenamefont
  {Ugander}}]{overgoor2019choosing}%
  \BibitemOpen
  \bibfield  {author} {\bibinfo {author} {\bibfnamefont {J.}~\bibnamefont
  {Overgoor}}, \bibinfo {author} {\bibfnamefont {A.}~\bibnamefont {Benson}},\
  and\ \bibinfo {author} {\bibfnamefont {J.}~\bibnamefont {Ugander}},\ }in\
  \href@noop {} {\emph {\bibinfo {booktitle} {The World Wide Web Conference}}}\
  (\bibinfo {organization} {ACM, New York},\ \bibinfo {year} {2019})\ pp.\
  \bibinfo {pages} {1409--1420}\BibitemShut {NoStop}%
\bibitem [{\citenamefont {Heider}(1958)}]{heider58}%
  \BibitemOpen
  \bibfield  {author} {\bibinfo {author} {\bibfnamefont {F.}~\bibnamefont
  {Heider}},\ }\href@noop {} {\emph {\bibinfo {title} {The Psychology of
  Interpersonal Relations}}}\ (\bibinfo  {publisher} {Psychology Press, Hove},\
  \bibinfo {year} {1958})\BibitemShut {NoStop}%
\bibitem [{\citenamefont {Holme}\ and\ \citenamefont
  {Newman}(2006)}]{holme2006nonequilibrium}%
  \BibitemOpen
  \bibfield  {author} {\bibinfo {author} {\bibfnamefont {P.}~\bibnamefont
  {Holme}}\ and\ \bibinfo {author} {\bibfnamefont {M.~E.~J.}\ \bibnamefont
  {Newman}},\ }\href@noop {} {\bibfield  {journal} {\bibinfo  {journal} {Phys.
  Rev. E}\ }\textbf {\bibinfo {volume} {74}},\ \bibinfo {pages} {056108}
  (\bibinfo {year} {2006})}\BibitemShut {NoStop}%
\bibitem [{\citenamefont {Evans}(2007)}]{evans2007exact}%
  \BibitemOpen
  \bibfield  {author} {\bibinfo {author} {\bibfnamefont {T.}~\bibnamefont
  {Evans}},\ }\href@noop {} {\bibfield  {journal} {\bibinfo  {journal} {Eur.
  Phys. J. B}\ }\textbf {\bibinfo {volume} {56}},\ \bibinfo {pages} {65}
  (\bibinfo {year} {2007})}\BibitemShut {NoStop}%
\bibitem [{\citenamefont {Vazquez}\ and\ \citenamefont
  {Egu{\'\i}luz}(2008)}]{vazquez2008analytical}%
  \BibitemOpen
  \bibfield  {author} {\bibinfo {author} {\bibfnamefont {F.}~\bibnamefont
  {Vazquez}}\ and\ \bibinfo {author} {\bibfnamefont {V.~M.}\ \bibnamefont
  {Egu{\'\i}luz}},\ }\href@noop {} {\bibfield  {journal} {\bibinfo  {journal}
  {New J. Phys.}\ }\textbf {\bibinfo {volume} {10}},\ \bibinfo {pages} {063011}
  (\bibinfo {year} {2008})}\BibitemShut {NoStop}%
\bibitem [{\citenamefont {Vazquez}\ \emph {et~al.}(2008)\citenamefont
  {Vazquez}, \citenamefont {Egu{\'\i}luz},\ and\ \citenamefont
  {Miguel}}]{vazquez2008generic}%
  \BibitemOpen
  \bibfield  {author} {\bibinfo {author} {\bibfnamefont {F.}~\bibnamefont
  {Vazquez}}, \bibinfo {author} {\bibfnamefont {V.~M.}\ \bibnamefont
  {Egu{\'\i}luz}},\ and\ \bibinfo {author} {\bibfnamefont {M.~S.}\ \bibnamefont
  {Miguel}},\ }\href@noop {} {\bibfield  {journal} {\bibinfo  {journal}
  {Phys.~Rev.~Lett.}\ }\textbf {\bibinfo {volume} {100}},\ \bibinfo {pages}
  {108702} (\bibinfo {year} {2008})}\BibitemShut {NoStop}%
\bibitem [{\citenamefont {Lindquist}\ \emph {et~al.}(2009)\citenamefont
  {Lindquist}, \citenamefont {Ma}, \citenamefont {Van~den Driessche},\ and\
  \citenamefont {Willeboordse}}]{lindquist2009network}%
  \BibitemOpen
  \bibfield  {author} {\bibinfo {author} {\bibfnamefont {J.}~\bibnamefont
  {Lindquist}}, \bibinfo {author} {\bibfnamefont {J.}~\bibnamefont {Ma}},
  \bibinfo {author} {\bibfnamefont {P.}~\bibnamefont {Van~den Driessche}},\
  and\ \bibinfo {author} {\bibfnamefont {F.~H.}\ \bibnamefont {Willeboordse}},\
  }\href@noop {} {\bibfield  {journal} {\bibinfo  {journal} {Physica D}\
  }\textbf {\bibinfo {volume} {238}},\ \bibinfo {pages} {370} (\bibinfo {year}
  {2009})}\BibitemShut {NoStop}%
\bibitem [{\citenamefont {Durrett}\ \emph {et~al.}(2012)\citenamefont
  {Durrett}, \citenamefont {Gleeson}, \citenamefont {Lloyd}, \citenamefont
  {Mucha}, \citenamefont {Shi}, \citenamefont {Sivakoff}, \citenamefont
  {Socolar},\ and\ \citenamefont {Varghese}}]{durrett2012graph}%
  \BibitemOpen
  \bibfield  {author} {\bibinfo {author} {\bibfnamefont {R.}~\bibnamefont
  {Durrett}}, \bibinfo {author} {\bibfnamefont {J.~P.}\ \bibnamefont
  {Gleeson}}, \bibinfo {author} {\bibfnamefont {A.~L.}\ \bibnamefont {Lloyd}},
  \bibinfo {author} {\bibfnamefont {P.~J.}\ \bibnamefont {Mucha}}, \bibinfo
  {author} {\bibfnamefont {F.}~\bibnamefont {Shi}}, \bibinfo {author}
  {\bibfnamefont {D.}~\bibnamefont {Sivakoff}}, \bibinfo {author}
  {\bibfnamefont {J.~E.~S.}\ \bibnamefont {Socolar}},\ and\ \bibinfo {author}
  {\bibfnamefont {C.}~\bibnamefont {Varghese}},\ }\href@noop {} {\bibfield
  {journal} {\bibinfo  {journal} {Proc. Natl. Acad. Sci. U.S.A.}\ }\textbf
  {\bibinfo {volume} {109}},\ \bibinfo {pages} {3682} (\bibinfo {year}
  {2012})}\BibitemShut {NoStop}%
\bibitem [{\citenamefont {Henry}\ \emph {et~al.}(2011)\citenamefont {Henry},
  \citenamefont {Pra{\l}at},\ and\ \citenamefont {Zhang}}]{henry2011emergence}%
  \BibitemOpen
  \bibfield  {author} {\bibinfo {author} {\bibfnamefont {A.~D.}\ \bibnamefont
  {Henry}}, \bibinfo {author} {\bibfnamefont {P.}~\bibnamefont {Pra{\l}at}},\
  and\ \bibinfo {author} {\bibfnamefont {C.}~\bibnamefont {Zhang}},\
  }\href@noop {} {\bibfield  {journal} {\bibinfo  {journal} {Proc. Natl. Acad.
  Sci. U.S.A.}\ }\textbf {\bibinfo {volume} {108}},\ \bibinfo {pages} {8605}
  (\bibinfo {year} {2011})}\BibitemShut {NoStop}%
\bibitem [{\citenamefont {Mobilia}(2015)}]{mobilia2015nonlinear}%
  \BibitemOpen
  \bibfield  {author} {\bibinfo {author} {\bibfnamefont {M.}~\bibnamefont
  {Mobilia}},\ }\href@noop {} {\bibfield  {journal} {\bibinfo  {journal}
  {Phys.~Rev.~E}\ }\textbf {\bibinfo {volume} {92}},\ \bibinfo {pages} {012803}
  (\bibinfo {year} {2015})}\BibitemShut {NoStop}%
\bibitem [{\citenamefont {Castellano}\ \emph
  {et~al.}(2009{\natexlab{b}})\citenamefont {Castellano}, \citenamefont
  {Mu{\~n}oz},\ and\ \citenamefont
  {Pastor-Satorras}}]{castellano2009nonlinear}%
  \BibitemOpen
  \bibfield  {author} {\bibinfo {author} {\bibfnamefont {C.}~\bibnamefont
  {Castellano}}, \bibinfo {author} {\bibfnamefont {M.~A.}\ \bibnamefont
  {Mu{\~n}oz}},\ and\ \bibinfo {author} {\bibfnamefont {R.}~\bibnamefont
  {Pastor-Satorras}},\ }\href@noop {} {\bibfield  {journal} {\bibinfo
  {journal} {Phys. Rev. E}\ }\textbf {\bibinfo {volume} {80}},\ \bibinfo
  {pages} {041129} (\bibinfo {year} {2009}{\natexlab{b}})}\BibitemShut
  {NoStop}%
\bibitem [{\citenamefont {Mellor}\ \emph {et~al.}(2017)\citenamefont {Mellor},
  \citenamefont {Mobilia},\ and\ \citenamefont
  {Zia}}]{mellor2017heterogeneous}%
  \BibitemOpen
  \bibfield  {author} {\bibinfo {author} {\bibfnamefont {A.}~\bibnamefont
  {Mellor}}, \bibinfo {author} {\bibfnamefont {M.}~\bibnamefont {Mobilia}},\
  and\ \bibinfo {author} {\bibfnamefont {R.~K.~P.}\ \bibnamefont {Zia}},\
  }\href@noop {} {\bibfield  {journal} {\bibinfo  {journal} {Phys. Rev. E}\
  }\textbf {\bibinfo {volume} {95}},\ \bibinfo {pages} {012104} (\bibinfo
  {year} {2017})}\BibitemShut {NoStop}%
\bibitem [{\citenamefont {Mobilia}\ \emph {et~al.}(2007)\citenamefont
  {Mobilia}, \citenamefont {Petersen},\ and\ \citenamefont
  {Redner}}]{mobilia2007role}%
  \BibitemOpen
  \bibfield  {author} {\bibinfo {author} {\bibfnamefont {M.}~\bibnamefont
  {Mobilia}}, \bibinfo {author} {\bibfnamefont {A.}~\bibnamefont {Petersen}},\
  and\ \bibinfo {author} {\bibfnamefont {S.}~\bibnamefont {Redner}},\
  }\href@noop {} {\bibfield  {journal} {\bibinfo  {journal} {J. Stat. Mech.:
  Theory Exp.}\ }\textbf {\bibinfo {volume} {2007}},\ \bibinfo {pages}
  {P08029}}\BibitemShut {NoStop}%
\bibitem [{\citenamefont {Liu}\ \emph {et~al.}(2013)\citenamefont {Liu},
  \citenamefont {Jolad}, \citenamefont {Schmittmann},\ and\ \citenamefont
  {Zia}}]{liu2013modeling}%
  \BibitemOpen
  \bibfield  {author} {\bibinfo {author} {\bibfnamefont {W.}~\bibnamefont
  {Liu}}, \bibinfo {author} {\bibfnamefont {S.}~\bibnamefont {Jolad}}, \bibinfo
  {author} {\bibfnamefont {B.}~\bibnamefont {Schmittmann}},\ and\ \bibinfo
  {author} {\bibfnamefont {R.~K.~P.}\ \bibnamefont {Zia}},\ }\href@noop {}
  {\bibfield  {journal} {\bibinfo  {journal} {J. Stat. Mech.: Theory Exp.}\
  }\textbf {\bibinfo {volume} {2013}},\ \bibinfo {pages} {P08001}}\BibitemShut
  {NoStop}%
\bibitem [{\citenamefont {Liu}\ \emph {et~al.}(2014)\citenamefont {Liu},
  \citenamefont {Schmittmann},\ and\ \citenamefont {Zia}}]{liu2014modeling}%
  \BibitemOpen
  \bibfield  {author} {\bibinfo {author} {\bibfnamefont {W.}~\bibnamefont
  {Liu}}, \bibinfo {author} {\bibfnamefont {B.}~\bibnamefont {Schmittmann}},\
  and\ \bibinfo {author} {\bibfnamefont {R.~K.~P.}\ \bibnamefont {Zia}},\
  }\href@noop {} {\bibfield  {journal} {\bibinfo  {journal} {J. Stat. Mech.:
  Theory Exp.}\ }\textbf {\bibinfo {volume} {2014}},\ \bibinfo {pages}
  {P05021}}\BibitemShut {NoStop}%
\bibitem [{\citenamefont {Bassler}\ \emph {et~al.}(2015)\citenamefont
  {Bassler}, \citenamefont {Dhar},\ and\ \citenamefont
  {Zia}}]{bassler2015networks}%
  \BibitemOpen
  \bibfield  {author} {\bibinfo {author} {\bibfnamefont {K.~E.}\ \bibnamefont
  {Bassler}}, \bibinfo {author} {\bibfnamefont {D.}~\bibnamefont {Dhar}},\ and\
  \bibinfo {author} {\bibfnamefont {R.~K.~P.}\ \bibnamefont {Zia}},\
  }\href@noop {} {\bibfield  {journal} {\bibinfo  {journal} {J. Stat. Mech.:
  Theory Exp.}\ }\textbf {\bibinfo {volume} {2015}},\ \bibinfo {pages}
  {P07013}}\BibitemShut {NoStop}%
\bibitem [{\citenamefont {Mobilia}(2013)}]{Mobilia-2013}%
  \BibitemOpen
  \bibfield  {author} {\bibinfo {author} {\bibfnamefont {M.}~\bibnamefont
  {Mobilia}},\ }\href@noop {} {\bibfield  {journal} {\bibinfo  {journal}
  {J.~Stat.~Phys.}\ }\textbf {\bibinfo {volume} {151}},\ \bibinfo {pages} {69}
  (\bibinfo {year} {2013})}\BibitemShut {NoStop}%
\bibitem [{\citenamefont {Mobilia}(2003)}]{Mobilia-2003}%
  \BibitemOpen
  \bibfield  {author} {\bibinfo {author} {\bibfnamefont {M.}~\bibnamefont
  {Mobilia}},\ }\href@noop {} {\bibfield  {journal} {\bibinfo  {journal}
  {Phys.~Rev.~Lett.}\ }\textbf {\bibinfo {volume} {91}},\ \bibinfo {pages}
  {028701} (\bibinfo {year} {2003})}\BibitemShut {NoStop}%
\bibitem [{\citenamefont {Galam}\ and\ \citenamefont
  {Jacobs}(2007)}]{Galam-2007}%
  \BibitemOpen
  \bibfield  {author} {\bibinfo {author} {\bibfnamefont {S.}~\bibnamefont
  {Galam}}\ and\ \bibinfo {author} {\bibfnamefont {F.}~\bibnamefont {Jacobs}},\
  }\href@noop {} {\bibfield  {journal} {\bibinfo  {journal} {Physica A}\
  }\textbf {\bibinfo {volume} {381}},\ \bibinfo {pages} {366} (\bibinfo {year}
  {2007})}\BibitemShut {NoStop}%
\bibitem [{\citenamefont {Sznajd-Weron}\ \emph {et~al.}(2011)\citenamefont
  {Sznajd-Weron}, \citenamefont {Tabiszewski},\ and\ \citenamefont
  {Timpanaro}}]{SW-2011}%
  \BibitemOpen
  \bibfield  {author} {\bibinfo {author} {\bibfnamefont {K.}~\bibnamefont
  {Sznajd-Weron}}, \bibinfo {author} {\bibfnamefont {M.}~\bibnamefont
  {Tabiszewski}},\ and\ \bibinfo {author} {\bibfnamefont {A.~M.}\ \bibnamefont
  {Timpanaro}},\ }\href@noop {} {\bibfield  {journal} {\bibinfo  {journal}
  {EPL}\ }\textbf {\bibinfo {volume} {96}},\ \bibinfo {pages} {48002} (\bibinfo
  {year} {2011})}\BibitemShut {NoStop}%
\bibitem [{\citenamefont {Acemoglu}\ \emph {et~al.}(2013)\citenamefont
  {Acemoglu}, \citenamefont {Como}, \citenamefont {Fagnani},\ and\
  \citenamefont {Ozdagla}}]{Acemoglu-2013}%
  \BibitemOpen
  \bibfield  {author} {\bibinfo {author} {\bibfnamefont {D.}~\bibnamefont
  {Acemoglu}}, \bibinfo {author} {\bibfnamefont {G.}~\bibnamefont {Como}},
  \bibinfo {author} {\bibfnamefont {F.}~\bibnamefont {Fagnani}},\ and\ \bibinfo
  {author} {\bibfnamefont {A.}~\bibnamefont {Ozdagla}},\ }\href@noop {}
  {\bibfield  {journal} {\bibinfo  {journal} {Math. Oper. Res.}\ }\textbf
  {\bibinfo {volume} {38}},\ \bibinfo {pages} {1} (\bibinfo {year}
  {2013})}\BibitemShut {NoStop}%
\bibitem [{\citenamefont {Liu}\ \emph {et~al.}(2012)\citenamefont {Liu},
  \citenamefont {Schmittmann},\ and\ \citenamefont
  {Zia}}]{liu2012extraordinary}%
  \BibitemOpen
  \bibfield  {author} {\bibinfo {author} {\bibfnamefont {W.}~\bibnamefont
  {Liu}}, \bibinfo {author} {\bibfnamefont {B.}~\bibnamefont {Schmittmann}},\
  and\ \bibinfo {author} {\bibfnamefont {R.~K.~P.}\ \bibnamefont {Zia}},\
  }\href@noop {} {\bibfield  {journal} {\bibinfo  {journal} {EPL}\ }\textbf
  {\bibinfo {volume} {100}},\ \bibinfo {pages} {66007} (\bibinfo {year}
  {2012})}\BibitemShut {NoStop}%
\bibitem [{\citenamefont {Prasetya}\ and\ \citenamefont
  {Murata}(2020)}]{prasetya2020model}%
  \BibitemOpen
  \bibfield  {author} {\bibinfo {author} {\bibfnamefont {H.~A.}\ \bibnamefont
  {Prasetya}}\ and\ \bibinfo {author} {\bibfnamefont {T.}~\bibnamefont
  {Murata}},\ }\href@noop {} {\bibfield  {journal} {\bibinfo  {journal}
  {Comput. Soc. Netw.}\ }\textbf {\bibinfo {volume} {7}},\ \bibinfo {pages} {2}
  (\bibinfo {year} {2020})}\BibitemShut {NoStop}%
\bibitem [{\citenamefont {Erd\"{o}s}\ and\ \citenamefont
  {R\'{e}nyi}(1959)}]{renyi1959random}%
  \BibitemOpen
  \bibfield  {author} {\bibinfo {author} {\bibfnamefont {P.}~\bibnamefont
  {Erd\"{o}s}}\ and\ \bibinfo {author} {\bibfnamefont {A.}~\bibnamefont
  {R\'{e}nyi}},\ }\href@noop {} {\bibfield  {journal} {\bibinfo  {journal}
  {Publicationes Mathematicate Debrecen}\ }\textbf {\bibinfo {volume} {6}},\
  \bibinfo {pages} {290} (\bibinfo {year} {1959})}\BibitemShut {NoStop}%
\bibitem [{\citenamefont {Li}\ \emph {et~al.}(2021)\citenamefont {Li},
  \citenamefont {Mobilia}, \citenamefont {Rucklidge},\ and\ \citenamefont
  {Zia}}]{companion}%
  \BibitemOpen
  \bibfield  {author} {\bibinfo {author} {\bibfnamefont {X.}~\bibnamefont
  {Li}}, \bibinfo {author} {\bibfnamefont {M.}~\bibnamefont {Mobilia}},
  \bibinfo {author} {\bibfnamefont {A.~M.}\ \bibnamefont {Rucklidge}},\ and\
  \bibinfo {author} {\bibfnamefont {R.~K.~P.}\ \bibnamefont {Zia}},\
  }\href@noop {} {\  (\bibinfo {year} {2021})},\ \Eprint
  {https://arxiv.org/abs/2107.13945} {e-print: arXiv:2107.13945
  [physics.soc-ph]} \BibitemShut {NoStop}%
\bibitem [{\citenamefont {Stewart}\ \emph {et~al.}(2019)\citenamefont
  {Stewart}, \citenamefont {Mosleh}, \citenamefont {Diakonova}, \citenamefont
  {Arechar}, \citenamefont {Rand},\ and\ \citenamefont
  {Plotkin}}]{stewart2019information}%
  \BibitemOpen
  \bibfield  {author} {\bibinfo {author} {\bibfnamefont {A.~J.}\ \bibnamefont
  {Stewart}}, \bibinfo {author} {\bibfnamefont {M.}~\bibnamefont {Mosleh}},
  \bibinfo {author} {\bibfnamefont {M.}~\bibnamefont {Diakonova}}, \bibinfo
  {author} {\bibfnamefont {A.~A.}\ \bibnamefont {Arechar}}, \bibinfo {author}
  {\bibfnamefont {D.~G.}\ \bibnamefont {Rand}},\ and\ \bibinfo {author}
  {\bibfnamefont {J.~B.}\ \bibnamefont {Plotkin}},\ }\href@noop {} {\bibfield
  {journal} {\bibinfo  {journal} {Nature}\ }\textbf {\bibinfo {volume} {573}},\
  \bibinfo {pages} {117} (\bibinfo {year} {2019})}\BibitemShut {NoStop}%
\bibitem [{\citenamefont {Nardini}\ \emph {et~al.}(2008)\citenamefont
  {Nardini}, \citenamefont {Kozma},\ and\ \citenamefont
  {Barrat}}]{nardini2008s}%
  \BibitemOpen
  \bibfield  {author} {\bibinfo {author} {\bibfnamefont {C.}~\bibnamefont
  {Nardini}}, \bibinfo {author} {\bibfnamefont {B.}~\bibnamefont {Kozma}},\
  and\ \bibinfo {author} {\bibfnamefont {A.}~\bibnamefont {Barrat}},\
  }\href@noop {} {\bibfield  {journal} {\bibinfo  {journal} {Phys. Rev. Lett.}\
  }\textbf {\bibinfo {volume} {100}},\ \bibinfo {pages} {158701} (\bibinfo
  {year} {2008})}\BibitemShut {NoStop}%
\bibitem [{\citenamefont {Iniguez}\ \emph {et~al.}(2009)\citenamefont
  {Iniguez}, \citenamefont {Kert{\'e}sz}, \citenamefont {Kaski},\ and\
  \citenamefont {Barrio}}]{iniguez2009opinion}%
  \BibitemOpen
  \bibfield  {author} {\bibinfo {author} {\bibfnamefont {G.}~\bibnamefont
  {Iniguez}}, \bibinfo {author} {\bibfnamefont {J.}~\bibnamefont
  {Kert{\'e}sz}}, \bibinfo {author} {\bibfnamefont {K.~K.}\ \bibnamefont
  {Kaski}},\ and\ \bibinfo {author} {\bibfnamefont {R.~A.}\ \bibnamefont
  {Barrio}},\ }\href@noop {} {\bibfield  {journal} {\bibinfo  {journal} {Phys.
  Rev. E}\ }\textbf {\bibinfo {volume} {80}},\ \bibinfo {pages} {066119}
  (\bibinfo {year} {2009})}\BibitemShut {NoStop}%
\bibitem [{\citenamefont {Gross}\ and\ \citenamefont
  {Blasius}(2008)}]{gross2008adaptive}%
  \BibitemOpen
  \bibfield  {author} {\bibinfo {author} {\bibfnamefont {T.}~\bibnamefont
  {Gross}}\ and\ \bibinfo {author} {\bibfnamefont {B.}~\bibnamefont
  {Blasius}},\ }\href@noop {} {\bibfield  {journal} {\bibinfo  {journal} {J. R.
  Soc. Interface}\ }\textbf {\bibinfo {volume} {5}},\ \bibinfo {pages} {259}
  (\bibinfo {year} {2008})}\BibitemShut {NoStop}%
\end{thebibliography}%
\end{document}